\documentclass[twocolumn,showpacs,preprintnumbers,amsmath,amssymb]{revtex4-1}

\usepackage{graphicx}
\usepackage{dcolumn}
\usepackage{bm}

\begin{document}

\preprint{}

\title{
$R$-parity violating two-loop level rainbowlike contribution\\
 to the fermion electric dipole moment}

\author{Nodoka~Yamanaka$^{1,2}$}
  \affiliation{$^1$Kobayashi-Maskawa Institute for the Origin of Particles and the Universe, Nagoya University, Furocho, Chikusa, Aichi 464-8602, Japan}
  \affiliation{$^2$Nishina Center for Accelerator-Based Science, RIKEN, Wako 351-0198, Japan}
  \email{nyamanaka@kmi.nagoya-u.ac.jp}

\date{\today}

\begin{abstract}
We analyze the two-loop level $R$-parity violating supersymmetric contribution to the electric and chromoelectric dipole moments of fermions with a lepton and a gaugino in the intermediate state.
It is found that this contribution can be sufficiently enhanced with large $\tan \beta$ and that it can have comparable size with the currently known $R$-parity violating Barr-Zee type process within TeV scale supersymmetry breaking.
We also give new upper limits on $R$-parity violating couplings given by the atomic electric dipole moment and molecular beam experiments.
\end{abstract}

\pacs{12.60.Jv, 11.30.Er, 13.40.Em, 14.80.Ly}
\maketitle

\section{\label{sec:intro}Introduction}

The standard model of particle physics is known to have many problems with observations, such as the matter abundance of our Universe, and many candidates of new physics have been invented so far.
Among them, the supersymmetry \cite{mssm,Haber:1984rc,Gunion:1984yn,Baer:2023cvi} is widely noticed, thanks to the possibility to resolve many phenomenological problems encountered in the standard model.

Recently, it has been suggested from the survey of metal-poor galaxies that the lepton-to-photon ratio is actually larger than the baryon one by six orders of magnitude \cite{Matsumoto:2022tlr}.
On the theoretical side, a consistent argument has been stated, that the sphaleron process which violates the baryon and lepton numbers $B+L$ is actually unphysical \cite{Yamanaka:2022vdt,Yamanaka:2022bfj}, so that $B$ and $L$ are indeed protected.
To realize the current Universe with the observed baryons and leptons, it is therefore required to conceive particle physics models with separate baryon and lepton number violations with local interactions.
Supersymmetric models with $R$-parity violation (RPV) \cite{Barbieri:1985ty,rpvphenomenology,Bhattacharyya:1997vv,Barbier:2004ez,Chemtob:2004xr,Dreiner:2023bvs} are indeed such good candidates.
The conservation of the $R$-parity is often assumed for keeping the lightest supersymmetric particle stable, but this is only an {\it ad hoc} assumption.
In the present case, RPV is rather welcome since it can fulfill all criteria of Sakharov, necessary for baryogenesis \cite{Sakharov:1967dj}.

To test the RPV, the electric dipole moment (EDM) \cite{He:1990qa,Bernreuther:1990jx,khriplovichbook,Ginges:2003qt,Pospelov:2005pr,Fukuyama:2012np,Hewett:2012ns,Engel:2013lsa,Yamanaka:2014mda,Yamanaka:2016umw,Yamanaka:2017mef,Chupp:2017rkp,Alarcon:2022ero} is a very suitable observable, since it is sensitive to the CP violation of new physics, while being extremely small in the standard model \cite{Kobayashi:1973fv,Czarnecki:1997bu,Seng:2014lea,Yamanaka:2015ncb,Yamanaka:2016fjj,Lee:2018flm,Yamaguchi:2020eub,Yamaguchi:2020dsy,Ema:2022yra}.
Here the advantage is that it can simultaneously probe the baryon or lepton number violating interactions and their CP phases, which are two important necessary conditions for baryogenesis according to Sakharov.
Due to the stringent experimental limits on the EDM measured in many systems \cite{Abel:2020gbr,Graner:2016ses,Caldwell:2022xwj,Roussy:2022cmp}, many $CP$ phases of new physics candidates, especially supersymmetric models \cite{Ellis:1982tk,Buchmuller:1982ye,Polchinski:1983zd,delAguila:1983dfr,Nanopoulos:1983xd,Dugan:1984qf,Nath:1991dn,Kizukuri:1991mb,Kizukuri:1992nj,Fischler:1992ha,Inui:1995fi,Ibrahim:1997gj,Ibrahim:1997nc,Ibrahim:1998je,Pokorski:1999hz,YaserAyazi:2006zw,YaserAyazi:2007xtv,Chang:1999zw,Chang:2002ex,Feng:2004vu,West:1993tk,Kadoyoshi:1996bc,Chang:1998uc,Pilaftsis:1999td,Pilaftsis:1999bq,Arkani-Hamed:2004zhs,Chang:2005ac,Giudice:2005rz,Feng:2006ei,Li:2008kz,Yamanaka:2012ia,Dai:1990xh,Arnowitt:1990je,Falk:1999tm,Brhlik:1998zn,Abel:2001vy,Pilaftsis:2002fe,Lebedev:2002ne,Demir:2002gg,Demir:2003js,Lebedev:2004va,Olive:2005ru,Degrassi:2005zd,Abel:2005er,Ellis:2008zy,Ellis:2010xm,Ellis:2011hp,Li:2010ax,Zhao:2013gqa,Dhuria:2013syh,Sala:2013osa,Hisano:2015rna,Nakai:2016atk,Cesarotti:2018huy,Yan:2020ocy,Hisano:2004tf,Endo:2003te,Cho:2004da,Hisano:2006mj,Hisano:2007cz,Hisano:2008hn,Altmannshofer:2010ad,Maekawa:2017xci,Yanase:2018qqq,Banerjee:2020zxw,Alvarado:2021qmn,Su:2022vju,Dao:2022rui,Kaneta:2023wrl}, were constrained so far.

\begin{figure}[htb]
\includegraphics[width=9cm]{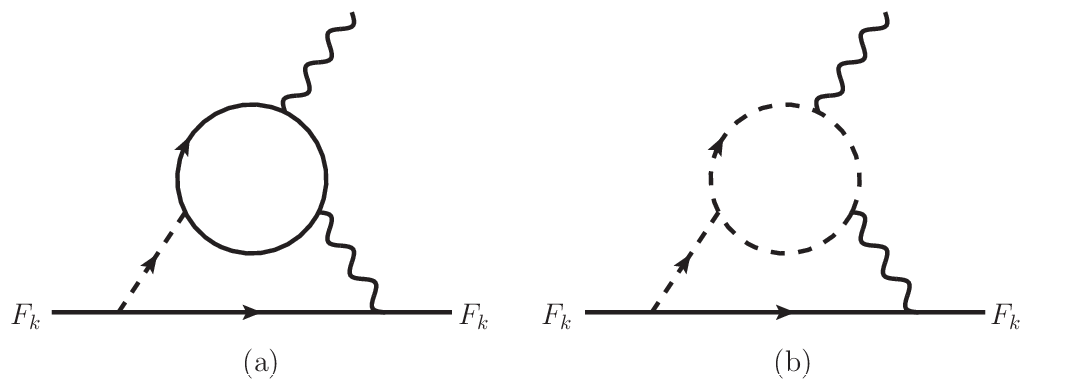}
\caption{\label{fig:barr-zee}
Examples of Barr-Zee type diagrams within RPV.
Thick, dashed, and wavy lines denote fermions, sfermions, and gauge bosons, respectively.
Figures (a) and (b) have fermion and sfermion inner loops, respectively.
The inner gauge bosons may be a photon, $W$ or $Z$ bosons for the EDM (with an external photon), and a gluon for the chromo-EDM (with an external gluon).
}
\end{figure}

\begin{figure}[htb]
\includegraphics[width=7cm]{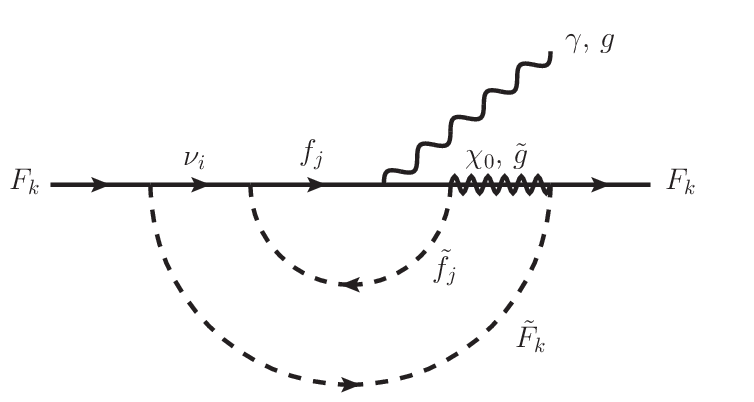}
\caption{\label{fig:RPV_rainbow} 
Example of a rainbow diagram contributing to the fermion EDM within RPV.
The neutralino is denoted by $\chi_0$ and the gluino by $\tilde g$.
}
\end{figure}

The RPV is also strongly constrained by EDM experiments, according to previous analyses up to two-loop level \cite{Godbole:1999ye,Abel:1999yz,Herczeg:1999me,Chang:2000wf,Choi:2000bg,Keum:2000ak,Kim:2001se,Faessler:2006vi,Faessler:2006at,Chiou:2006qk,Yamanaka:2012hm,Yamanaka:2012zy,Yamanaka:2012zq,Yamanaka:2012ep,Yamanaka:2014mda,Yamanaka:2014nba,Yamanaka:2012zq,Yanase:2018qqq}.
It is currently known that the leading contribution is given by the Barr-Zee type diagrams \cite{Barr:1990vd} generated by trilinear RPV interactions (see Fig. \ref{fig:barr-zee}) \cite{Yamanaka:2012hm,Yamanaka:2012zq,Yamanaka:2014mda}, under the assumption of the absence of soft supersymmetry breaking in the RPV sector.
The Barr-Zee type process is also relevant in the R-parity conserving sector, and it often provides a large contribution, but it was pointed that the two-loop level {\it rainbowlike} contribution may have a comparable size as regard the fermion EDM \cite{Pilaftsis:1999bq,Yamanaka:2012ia} and the muon anomalous magnetic moment \cite{Fargnoli:2013zda,Fargnoli:2013zia}.
This contribution has a fermion-sfermion inner loop, connected to the fermion external line with a Higgsino and a gaugino, and the flavor structure as well as the coupling constants involved  are exactly the same as for the Barr-Zee type diagrams.
In the RPV sector, similar rainbowlike diagrams can also be drawn by replacing the Higgsino by a neutrino (see Fig. \ref{fig:RPV_rainbow}), and it is therefore of necessity to evaluate and analyze this potentially sizable contribution.
In this paper, we therefore quantify the rainbowlike diagrams in RPV, and update the constraints on it using the latest results of molecular beam and atomic EDM experiments.

This paper is organized as follows.
We first introduce the supersymmetric and RPV interactions relevant in this discussion in the next section.
In Section \ref{sec:rpvedm}, we review the Barr-Zee type diagrams within RPV and calculate the RPV rainbowlike contribution to the fermion EDM and to the chromo-EDM.
In Section \ref{sec:many-body}, we provide formulae which relate the elementary level CP-violating processes to observables of molecular beam and atomic EDM experiments.
We then analyze in Section \ref{sec:analysis} the effect of the RPV supersymmetric rainbow diagrams by comparing with the EDM and the chromo-EDM generated by the RPV Barr-Zee type diagrams, and set upper limits on RPV couplings from experimental data.
The last section is devoted to the summary.
We have also collected the detailed derivation of important formulae in the Appendices.

\section{\label{sec:lagrangian}RPV supersymmetric interactions}

Let us first give the Lagrangian of the particles relevant in this discussion.
The RPV interactions relevant in this discussion are given by the following superpotential:
\begin{eqnarray}
W_{ R\hspace{-.5em}/} &=& \frac{1}{2} \lambda_{ijk} \epsilon_{ab}
 L_i^a L_j^b (E^c)_k
 +\lambda'_{ijk} \epsilon_{ab} L_i^a Q_j^b ( D^c)_k \nonumber\\
&& + \frac{1}{2} \lambda''_{ijk} (U^c)_i (D^c)_j (D^c)_k \ ,
\label{eq:superpotential}
\end{eqnarray}
with $i,j,k=1,2,3$ indicating the generation, $a,b=1,2$ the $SU(2)_L$ indices. 
The $SU(3)_c$ indices have been omitted. 
Here $L$, $E^c$, $Q$, $U^c$ and $D^c$ denote the lepton doublet, charged lepton singlet, quark doublet, up-type quark singlet, and down-type quark singlet left-chiral superfields, respectively.
The bilinear term has been omitted in our discussion. 

We will from now only consider the lepton number violating RPV interactions since, as mentioned in the previous section, the observation currently suggests that the lepton number asymmetry is very large \cite{Matsumoto:2022tlr}.
Moreover, baryon number violating RPV interactions (often denoted as $\lambda''_{ijk}$) lead to rapid proton and nuclear decays, which are constrained by experiments.
This RPV superpotential yields the following lepton number violating Yukawa interactions:
\begin{eqnarray}
{\cal L }_{ R\hspace{-.5em}/\,} &=&
- \frac{1}{2} \lambda_{ijk} \left[
\tilde \nu_i \bar e_k P_L e_j +\tilde e_{Lj} \bar e_k P_L \nu_i + \tilde e_{Rk}^\dagger \bar \nu_i^c P_L e_j \right.\nonumber\\
&&\hspace{4.5em} -(i \leftrightarrow j ) \Bigr] 
\nonumber\\
&&-\lambda'_{ijk} \left[
\tilde \nu_i \bar d_k P_L d_j + \tilde d_{Lj} \bar d_k P_L \nu_i +\tilde d_{Rk}^\dagger \bar \nu_i^c P_L d_j \right. \nonumber\\
&&\hspace{3.5em} \left. -\tilde e_{Li} \bar d_k P_L u_j - \tilde u_{Lj} \bar d_k P_L e_i - \tilde d_{Rk}^\dagger \bar e_i^c P_L u_j \right] \nonumber\\
&&\hspace{4em}  + ({\rm h.c.})  \ .
\end{eqnarray}
The soft breaking terms of the RPV sector will not be considered in this discussion.

We should also give the supersymmetric interactions of the $R$-parity conserving sector.
The neutralino mass matrix is given as follows:
\begin{equation}
{\cal L}_{\chi_0} = -\frac{1}{2} \bar \chi_{0R} M_N \chi_{0L} + {\rm H.c.}\, ,
\end{equation}
with
\begin{equation}
M_N
=
\left(
\begin{array}{cccc}
0&\mu e^{i\theta_\mu} & \frac{-e v_u}{\sqrt{2} \sin \theta_W} & \frac{e v_u}{\sqrt{2} \cos \theta_W} \cr
\mu e^{i\theta_\mu} &0 & \frac{e v_d}{\sqrt{2} \sin \theta_W} & \frac{-e v_d}{\sqrt{2} \cos \theta_W} \cr
\frac{-e v_u}{\sqrt{2} \sin \theta_W} & \frac{e v_d}{\sqrt{2} \sin \theta_W} &m_{\lambda_1}  e^{i\theta_1}&0\cr
\frac{e v_u}{\sqrt{2} \cos \theta_W} & \frac{-e v_d}{\sqrt{2} \cos \theta_W} &0&m_{\lambda_2} e^{i\theta_2}\cr
\end{array}
\right)\, ,
\label{eq:neutralinomassmatrix}
\end{equation}
where the corresponding neutralino field vector $\chi_0$ has as the first two components the up- and down-type Higgsinos, and the last two components refer to the $U(1)_Y$ and the $SU(2)_L$ gauginos.
Here $\theta_W$ is the Weinberg angle, $v_u = v/\sqrt{1+\cot^2 \beta}$ and $v_d = v/\sqrt{1+\tan^2 \beta}$ where $v\approx 246$ GeV is the vacuum expectation value of the Higgs field. 
Similarly, the chargino Lagrangian is given by
\begin{equation}
{\cal L}_{\chi_\pm} 
= 
-\bar \chi_{R} M_C \chi_{L}
-\bar \chi_{L} M_C^\dagger \chi_{R}
\, ,
\end{equation}
with
\begin{equation}
M_C
=
\left(
\begin{array}{cccc}
-\mu e^{i\theta_\mu} & \frac{-e v_u}{\sqrt{2} \sin \theta_W} \cr
\frac{-e v_d}{\sqrt{2} \sin \theta_W} & m_{\lambda_2} e^{i\theta_2} \cr
\end{array}
\right)\, ,
\label{eq:charginomassmatrix}
\end{equation}
where the upper (lower) component denotes the charged Higgsino ($SU(2)_L$ gaugino).
In this work, we assume that the supersymmetry breaking scale is beyond TeV, which is strongly suggested by the recent results of LHC experiments \cite{CMS:2023ktc,CMS:2023yzg,ATLAS:2023lfr,ATLAS:2023afl}, so we neglect the mixing between gauginos and Higgsinos which always take a factor of Higgs vacuum expectation value ($v_u$ or $v_d$). 
The transition to down-type Higgsinos involves an additional suppression by $1/\tan \beta$, which is actually the case of the present work with RPV interactions.
In our discussion, the Higgsinos are thus irrelevant.
The components of the mass matrix have $CP$ phases in the general case, and they can manifest themselves as observable effects when mass insertions occur in the process.
For the gluino, we remove its $CP$ phase as usual.

The sfermion mass matrix is given by
\begin{equation}
M_{\tilde f}^2
\approx
\left(
\begin{array}{cc}
m_{\tilde f_L}^2 & m_f (A_f^* - R_f \mu e^{i\theta_\mu}) \\
m_f (A_f - R_f \mu e^{-i\theta_\mu} ) & m_{\tilde f_R}^2 \\
\end{array}
\right)\ ,
\end{equation}
where $R_f = \cot \beta$ for up-type squarks, and $R_f = \tan \beta$ for down-type squarks and charged sleptons.
The trilinear soft supersymmetry breaking couplings $A_f$ are assumed to be proportional to the quark or lepton mass matrices and they may have CP phases.
Note that in this discussion, only the down-type squarks and the charged sleptons are relevant so enhancements due to $\tan \beta$ may occur.
To obtain the sfermion mass eigenbasis, we introduce the following unitary matrix with the sfermion mixing angle $\theta_f$  and the $CP$ phase $\delta_f$:
\begin{equation}
\left(
\begin{array}{c}
\tilde f_L \\
\tilde f_R \\
\end{array}
\right)
= 
\left(
\begin{array}{cc}
1&0 \\
0& e^{i\delta_f} \\
\end{array}
\right)
\left(
\begin{array}{cc}
\cos \theta_f & \sin \theta_f \\
-\sin \theta_f & \cos \theta_f \\
\end{array}
\right)
\left(
\begin{array}{c}
\tilde f_1 \\
\tilde f_2 \\
\end{array}
\right) \ ,
\end{equation}
with $\cos \theta_f \approx \frac{2 m_f R_f \mu}{m_{\tilde f_L}^2 - m_{\tilde f_R}^2}$ and $\sin \theta_f \approx 1$ for $m_f |A_f| \ll {m_f R_f \mu} \ll {m_{\tilde f_L}^2 - m_{\tilde f_R}^2}$.
In this approximation, the two mass eigenvalues are just $m_{\tilde f_1} \approx m_{\tilde f_R}$ and $m_{\tilde f_2} \approx m_{\tilde f_L}$.
The above inequality fails when $m_{\tilde f_L}$ and $m_{\tilde f_R}$ are degenerate, but a mass splitting of 10\% is enough to marginally fulfill it.
In this discussion, we use the overall $CP$ phase $\delta_f = \arg (A_{f} - R_f \mu e^{-i\theta_\mu})$.

We should also define the gaugino-sfermion-fermion interactions.
They are given by
\begin{eqnarray}
{\cal L}_{\lambda}
&=&
\sum_{\tilde f} \sqrt{2} g^{(n)}_{\tilde f_{L/R}} \tilde f_{L/R}^\dagger \bar \lambda_{n} P_{L/R} f 
\nonumber\\
&&
+\sum_{\tilde f} \frac{e}{\sin \theta_W} \tilde f_{dL}^\dagger \bar \lambda'_{2} P_{L} f_u 
\nonumber\\
&&
+\sum_{\tilde f} \frac{e}{\sin \theta_W} \tilde f_{uL}^\dagger \bar {\lambda'_2}^c P_{L} f_d 
+{\rm (h.c.)}
\ .
\label{eq:gaugino-int}
\end{eqnarray}
The convention for the sign of the gauge coupling is $D^\mu \equiv \partial^\mu -ig A_a^\mu t_a$ where $t_a$ is the generator of the gauge group.
The index $n$ denotes the gauge group of the gaugino $\lambda_n$.
We define $\lambda_1$ as the $U(1)_Y$ gaugino, $\lambda_2$ as the neutral $SU(2)_L$ gaugino, $\lambda_3 = \tilde g$ as the gluino, and $\lambda'_2$ is the charged $SU(2)_L$ gaugino (=chargino). 
The fermion-sfermion-gaugino coupling constants are given as follows.
for the gluino couplings, we have $g^{(3)}_{\tilde q_L} = \frac{1}{2} g_s$ and $g^{(3)}_{\tilde q_R} = -\frac{1}{2} g_s$ with the QCD coupling $g_s = \sqrt{4 \pi \alpha_s}$.
For the couplings of $\lambda_1$, we have $g^{(1)}_{\tilde q_L} = \frac{1}{6}\frac{e}{\cos \theta_W}$, $g^{(1)}_{\tilde d_R} = \frac{1}{3}\frac{e}{\cos \theta_W}$, $g^{(1)}_{\tilde l_L} = -\frac{1}{2}\frac{e}{\cos \theta_W}$, and $g^{(1)}_{\tilde e_R} = \frac{e}{\cos \theta_W}$.
We have finally for the $\lambda_2$ couplings $g^{(2)}_{\tilde d_L} = -\frac{1}{2}\frac{e}{\sin \theta_W}$ and $g^{(2)}_{\tilde l_L} = -\frac{1}{2}\frac{e}{\sin \theta_W}$.

\section{\label{sec:rpvedm}R-parity violating contribution to the EDM}

\subsection{\label{sec:rpvbarr-zee}Barr-Zee type diagrams and four-quark interaction}

The leading order RPV contribution to the EDM of electrons and atoms is given by the fermion EDM interaction
\begin{equation}
{\cal L}_{\rm EDM}
=
- \frac{i}{2} \sum_F d_F \bar F \sigma_{\mu \nu} F^{\mu \nu} \gamma_5 F
,
\end{equation}
where $F$ is a down-type quark or charged lepton, by the chromo-EDM of the down-type quark $D$
\begin{eqnarray}
{\cal L}_{\rm cEDM}
&=&
- \frac{i}{2} \sum_{D} d^c_D g_s \bar D \sigma_{\mu \nu} G_a^{\mu \nu} t_a \gamma_5 D
,
\end{eqnarray}
and by the CP-odd four-fermion interactions between two fermions $f$ and $F$
\begin{equation}
{\cal L}_{\rm 4f}
=
\sum_{f,F} C_{fF} \bar f f \, \bar F i \gamma_5 F
.
\label{eq:4fint}
\end{equation}
We emphasize that the fermions are either charged leptons or down-type quarks in this RPV analysis.

The four-fermion interaction is generated at the tree level in RPV supersymmetry \cite{Herczeg:1999me,Faessler:2006vi,Faessler:2006at,Yamanaka:2014nba,Yanase:2018qqq}.
Its Lagrangian is just
\begin{equation}
{\cal L}_{\rm 4f}
=
\frac{{\rm Im}(\hat{\lambda}_{ijj} \tilde \lambda^*_{ikk})}{m_{\tilde \nu_i}^2}
[\bar f_j f_j \cdot \bar F_k i\gamma_5 F_k 
-\bar f_j i\gamma_5 f_j \cdot \bar F_k F_k]
,
\label{eq:RPV4f}
\end{equation}
where $\hat \lambda$ and $\tilde \lambda$ are equal to $\lambda$ or $\lambda'$, depending on whether $f$ and $F$ are a lepton or a quark, respectively.

Regarding the EDM and the chromo-EDM, the leading order contribution is generated by the trilinear RPV interaction and arises at the two-loop level \cite{Godbole:1999ye,Abel:1999yz,Chang:2000wf,Yamanaka:2014mda}.
At this order, the Barr-Zee type diagram (see Fig. \ref{fig:barr-zee}) is known to be large.
They are approximately given by
\begin{eqnarray}
d^{\rm BZ}_{F_k}
&\approx &
d^{\gamma}_{F_k}
+
d^{Z}_{F_k}
+
d^{W}_{F_k}
,
\label{eq:BZtot}
\end{eqnarray}
where the inner photon + fermion loop diagram is expressed as \cite{Yamanaka:2012hm,Yamanaka:2014mda}
\begin{equation}
d^{\gamma}_{F_k}
\approx 
{\rm Im } ( \hat \lambda_{ijj} \tilde \lambda^*_{ikk}) \frac{n_c Q_f^2 Q_F e \alpha_{\rm em} m_{f_j}}{16\pi^3 m_{\tilde \nu_i }^2 } 
\left( 2 + \ln \frac{m_{f_j}^2 }{m_{\tilde \nu_i }^2} \right) 
.
\label{eq:Barr-Zee}
\end{equation}
This approximation holds for $m_{\tilde \nu_i } \gg m_f$.
The inner $Z$ boson + fermion loop diagram is given by
\begin{equation}
d^{Z}_{F_k}
\approx 
{\rm Im } ( \hat \lambda_{ijj} \tilde \lambda^*_{ikk}) \frac{n_c Q_f \alpha_f \alpha_F e \alpha_{\rm em} m_{f_j}}{16\pi^3 m_{\tilde \nu_i }^2 } 
\ln \frac{m_Z^2 }{m_{\tilde \nu_i }^2}
,
\label{eq:ZBZ}
\end{equation}
where $\alpha_{f} = \alpha_d \equiv \frac{1}{12} (\tan \theta_W - 3 \cot \theta_W) \approx -0.42 \,(\alpha_{f} = \alpha_l \equiv \frac{1}{4} (3 \tan \theta_W - \cot \theta_W) \approx -0.065)$ when $f$ is a down-type quark (charged lepton).
This contribution is smaller than 10\% of $d^{\gamma}_{F_k}$, and it is thus not important.
Finally, the contribution of the inner $W$ boson + fermion loop diagrams \cite{Yamanaka:2012ep} is
\begin{eqnarray}
d^{W}_{F_k}
&\approx &
{\rm Im } ( \hat \lambda_{iaj} \tilde \lambda^*_{ibk}) \frac{s_f e \alpha_{\rm em} V_{aj} V_{bk} m_{f_j}}{128\pi^3 \sin^2 \theta_W m_{\tilde e_{Li} }^2}
\nonumber\\
&& \times
\Biggl[
\left( \frac{3 m_{f'_a}^4}{m_W^4}-\frac{2 m_{f'_a}^2}{m_W^2}  \right) 
\left\{ {\rm Li}_2 \left( 1- \frac{m_W^2}{m_{f'_a}^2}  \right) - \frac{\pi^2}{6} \right\}
\nonumber\\
&& \hspace{2em}
+ \frac{3 m_{f'_a}^2}{m_W^2} \left(1-\ln \frac{m_W^2}{ m_{f'_a}^2} \right)
+\frac{1}{2}\ln \frac{m_W^2}{ m_{\tilde e_{Li} }^2}
\Biggr]
,
\label{eq:WBZ}
\end{eqnarray}
where $f'$ is an up-type quark (neutrino), $s_f = +1$ ($-1$), and $V$ is the Cabibbo-Kobayashi-Maskawa matrix (unit matrix) element if $f$ of the inner loop is a down-type quark (charged lepton). 
This approximation assumes $m_{\tilde e_{Li} }^2 \gg m_W^2 , m_{u_a}^2$.
In this work, we only consider the flavor diagonal case, i.e. $i=a$ and $b=k$.
The $W$ boson contribution is not small, so we have to take it into account.

The quark chromo-EDM has an expression similar to the inner photon diagram contribution:
\begin{eqnarray}
d^{c;{\rm BZ}}_{D_k} 
&\approx& 
{\rm Im} (\lambda'_{ijj} \lambda'^*_{ikk}) \frac{\alpha_{\rm s} m_{d_j}}{32\pi^3 m_{\tilde \nu_i }^2} 
\left( 2 + \ln \frac{m_{d_j}^2 }{m_{\tilde \nu_i }^2} \right) 
.
\label{eq:chromo-Barr-Zee}
\end{eqnarray}
Here the fermions $D$ and $d$ are down-type quarks.

So far, we have only seen the inner fermion loop contribution [diagram of Fig. \ref{fig:barr-zee} (a)].
We have actually neglected the inner sfermion loop diagrams [an example is shown in Fig. \ref{fig:barr-zee} (b)] since they are small in TeV scale supersymmetry breaking \cite{Yamanaka:2012zq}.

\subsection{\label{sec:rainbow}Rainbowlike diagrams}

\begin{figure*}[htb]
\begin{center}
\includegraphics[width=18cm]{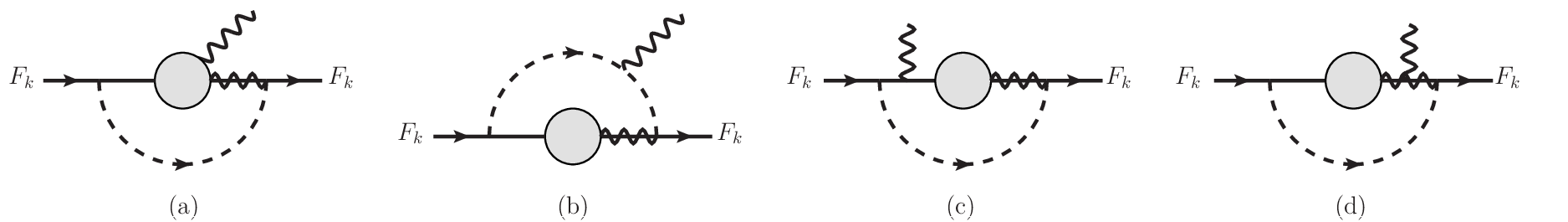}
\caption{\label{fig:rainbow_type}
Possible insertions of one-loop effective vertices (grey blobs) for the rainbow diagrams in RPV supersymmetry.
The dashed, solid and solid-wavy lines denote the sfermions, leptons and gauginos, respectively.
The external wavy line is a photon for the EDM, and a gluon for the chromo-EDM.
The diagrams (c) and (d) only receive contribution from the intermediate chargino and charged lepton.
}
\end{center}
\end{figure*}

Let us now calculate the RPV rainbow diagrams.
We can split the rainbow diagrams contributing to the EDM into two parts, one having a neutralino and a neutrino in the intermediate state, and another one having a chargino and a charged lepton, as
\begin{equation}
d_{F}^{\rm rb}
=
d^{\chi_0}_{F}
+d^{\chi_-}_{F}
.
\end{equation}
The rainbow contribution with a neutralino and a neutrino, $d^{\chi_0}_{F}$, comprises two types of diagrams.
The first type is the insertion of the effective one-loop level neutrino-gaugino-gauge boson vertex [Fig. \ref{fig:rainbow_type} (a)].
The second one is the insertion of the effective one-loop level neutrino-gaugino transition [Fig. \ref{fig:rainbow_type} (b)].
The calculation of these RPV rainbow diagrams is very similar to that in the minimal supersymmetric standard model with R-parity conservation \cite{Yamanaka:2012ia}.
The rainbow contribution with a chargino and a charged lepton, $d^{\chi_-}_{F}$, involves additional types of diagrams where the external photon is attached to these charge-one particles [Figs. \ref{fig:rainbow_type} (c) and (d)].

The final EDM is obtained by calculating the rainbow diagrams up to the first order in the external momentum carried by the gauge boson (recall that the EDM is the first order coefficient of the multipolar expansion of the CP-odd form factor).
In this work, we neglect the fermion mass insertions which are much smaller than those of sparticles.
The neutralino contribution to the EDM of a fermion $F_k$ ($k$ is the generation number) is given by
\begin{widetext}
\begin{eqnarray}
d^{\chi_0}_{F_k}
&=&
\sum_{n=1,2} 
\sum_{\tilde F= \tilde F_L , \tilde F_R} 
{\rm Im} ( \hat \lambda_{ijj} \tilde \lambda^*_{ikk} e^{i(\theta_n -\delta_{f_j})})
\frac{n_c e }{256 \pi^4} m_{\lambda_n } \sin \theta_{f_j} \cos \theta_{f_j} 
g^{(n)}_{\tilde F} 
\nonumber\\
&&
\times
\Bigl[
Q_f \Bigl( g^{(n)}_{\tilde f_L} -g^{(n)}_{\tilde f_R} \Bigr)
s_{\tilde F}\, 
G'(m_{\lambda_n}^2, 0 , m_{\tilde F_{k}}^2 , m_{\tilde f_{1j}}^2 \, )
-
Q_F \Bigl( g^{(n)}_{\tilde f_L} +g^{(n)}_{\tilde f_R} \Bigr)
m_{\tilde F_{k}}^2
G''(m_{\lambda_n}^2, 0 , m_{\tilde F_{k}}^2 , m_{\tilde f_{1j}}^2 \, )
\Bigr]
\nonumber\\
&&
-(m_{\tilde f_{1j}}^2 \leftrightarrow m_{\tilde f_{2j}}^2)
,
\label{eq:edmnug}
\end{eqnarray}
where the first and second terms of the square brackets correspond to the diagrams of Fig. \ref{fig:rainbow_type} (a) and (b), respectively.
We note that in the leading order RPV, the fermions (sfermions) $F$ and $f$ ($\tilde F$ and $\tilde f$) are down-type quarks or charged leptons (squarks or charged sleptons).
Here $i$ and $j$ are the flavor indices of the intermediate neutrino and the fermion-sfermion pair of the inner loop, respectively, and $n_c=3$ (=1) with an inner quark-squark (lepton-slepton) loop.
The RPV couplings are given by $\hat \lambda = \lambda$ ($= \lambda'$) for the charged lepton-slepton (quark-squark) inner loop, and $\tilde \lambda = \lambda$ ($= \lambda'$) for the external lepton (quark), in the same way as defined below Eq. (\ref{eq:RPV4f}).
The constant $s_{\tilde F}$ is $+1$ for the left-handed sfermion $\tilde F_{Lk}$ and $-1$ for the right-handed sfermion $\tilde F_{Rk}$.
The electric charges of the inner loop fermion (sfermion) and the external fermion in the unit of $e$ are $Q_f$ and $Q_F$, respectively.
The functions $G'$ and $G''$ are defined by
\begin{eqnarray}
G'(a,b,c,d) 
&\equiv &
\frac{a \ln a -d {\rm Li}_2 \left( 1- \frac{a}{d}\right) }{(a-b)(a-c)}
+\frac{b \ln b -d {\rm Li}_2 \left( 1- \frac{b}{d}\right) }{(b-a)(b-c)}
+\frac{c \ln c -d {\rm Li}_2 \left( 1- \frac{c}{d}\right) }{(c-a)(c-b)}
\, ,
\label{eq:F'}
\\
G''(a,b,c,d) 
&\equiv &
\frac{(a-d)^2}{(a-b)(a-c)^3} \left[ {\rm Li}_2 \left( 1-\frac{a}{d} \right) -{\rm Li}_2 \left( 1-\frac{c}{d} \right) \right]
-\frac{(b-d)^2}{(a-b)(b-c)^3} \left[ {\rm Li}_2 \left( 1-\frac{b}{d} \right) - {\rm Li}_2 \left( 1-\frac{c}{d} \right) \right]
\nonumber\\
&&
-\frac{ a^2 \ln d +ad}{(a-b)(a-c)^3} \ln \frac{a}{c}
+\frac{ b^2 \ln d +bd}{(a-b)(b-c)^3}  \ln \frac{b}{c}
+\frac{d}{a-b} \left[ 
\frac{1}{(a-c)^2 } 
-\frac{1}{ (b-c)^2} 
\right] \cdot \left[ 1+ \ln \frac{d}{c} \right]
\, ,
\label{eq:F''}
\end{eqnarray}
with ${\rm Li}_2 (x)$ denoting the dilogarithm function.
In Ref. \cite{Yamanaka:2012ia}, the notations used to define the functions $G'$ and $G''$ were $F'$ and $F''$, but we changed them to $G'$ and $G''$ so as to avoid confusion with fermions used in this work.
The zero arguments of Eq. (\ref{eq:edmnug}) are due to massless neutrinos.
The detail of the computation of the contributions of diagrams (a) and (b) of Fig. \ref{fig:rainbow_type} to $d^{\chi_0}_{F}$ is presented in Appendices \ref{sec:type1rainbow} and \ref{sec:type2rainbow}, respectively.

The EDM of the down-type quark or of the charged lepton generated by the rainbow diagrams with chargino and charged lepton is given by
\begin{eqnarray}
d^{\chi_-}_{F_k}
&=&
{\rm Im} \bigl[ \hat \lambda_{ijj} \tilde \lambda^*_{ikk} e^{i(\theta_2 -\delta_{f_j} )} \bigr]
\frac{ e \alpha_{\rm em} n_c V_{jj} V_{kk}}{128 \pi^3 \sin^2 \theta_W }  \sin \theta_{f_j} \cos \theta_{f_j} m_{\lambda_2} 
\nonumber\\
&& 
\times \Biggl\{
Q_{f'} G' ( m_{\lambda_2}^2 , m_{e_i}^2 , m_{\tilde F'_{Lk}}^2 , m_{\tilde f_{1j}}^2 )
- Q_{F'} m_{\tilde F'_{Lk}}^2 G'' ( m_{\lambda_2}^2 , m_{e_i}^2 , m_{\tilde F'_{Lk}}^2 , m_{\tilde f_{1j}}^2 )
+ m_{\tilde F'_{Lk}}^2 G^{c_1} ( m_{\lambda_2}^2 , m_{e_i}^2 , m_{\tilde F'_Lk}^2 , m_{\tilde f_{1j}}^2 )
\nonumber\\
&&  \hspace{2em}
- G^{c_2} ( m_{e_i}^2 , m_{\tilde F'_{Lk}}^2 , m_{\lambda_2}^2 , m_{\tilde f_{1j}}^2 )
- m_{\tilde F'_{Lk}}^2 
\biggl[
3G^{c_3} ( m_{\lambda_2}^2 , m_{e_i}^2 , m_{\tilde F'_{Lk}}^2 , m_{\tilde f_{1j}}^2 )
+ 
2m_{\lambda_2}^2 G^{c_4} ( m_{\lambda_2}^2 , m_{e_i}^2 , m_{\tilde F'_{Lk}}^2 , m_{\tilde f_{1j}}^2 )
\biggr]
\Biggr\}
\nonumber\\
&&
-(m_{\tilde f_{1j}}^2 \leftrightarrow m_{\tilde f_{2j}}^2)
,\ \ \ \ \ 
\label{eq:chargino_rainbow}
\end{eqnarray}
where we neglected higher order terms in the charged lepton mass $m_{e_i}$.
Here $f',F'$ ($\tilde f',\tilde F'$) are up-type quarks (squarks) or neutrinos (sneutrinos) respecting the definition of $\hat \lambda$ and $\tilde \lambda$ as given above.
Their electric charges $Q_{f'}$ and $Q_{F'}$ are also defined accordingly (+2/3 for the up-type quarks/squarks and zero for the neutrinos/sneutrinos).
The functions $G^c$'s used in the formula above are defined as
\begin{eqnarray}
G^{c_1} (a,b,c,d) 
&\equiv &
\frac{a -d}{(a-b)(a-c)^2} \left[ {\rm Li}_2 \left( 1-\frac{a}{d} \right) -{\rm Li}_2 \left( 1-\frac{c}{d} \right) \right] 
+\frac{b-d}{(b-a)(b-c)^2} \left[ {\rm Li}_2 \left( 1-\frac{b}{d} \right) -{\rm Li}_2 \left( 1-\frac{c}{d} \right) \right] 
\nonumber\\
&&
+\frac{a }{(a-b)(a-c)^2} \ln d \ln \frac{c}{a} 
+\frac{b}{(b-a)(b-c)^2} \ln d \ln \frac{c}{b} 
\, ,
\label{eq:fc1}
\\
G^{c_2} (a,b,c,d) 
&\equiv &
-\frac{(a-d)^2}{2(a-b)(a-c)^2} \left[ {\rm Li}_2 \left( 1-\frac{a}{d} \right) -{\rm Li}_2 \left( 1-\frac{c}{d} \right) \right] 
-\frac{(b-d)^2}{2(b-a)(b-c)^2} \left[ {\rm Li}_2 \left( 1-\frac{b}{d} \right) -{\rm Li}_2 \left( 1-\frac{c}{d} \right) \right] 
\nonumber\\
&& 
+\frac{(c-d)\ln c +d ( 1+ \ln d )}{2(c-a)(c-b)} 
-\frac{ad + a^2 \ln d}{2(a-b)(a-c)^2} \ln \frac{c}{a} 
-\frac{bd + b^2 \ln d}{2(b-a)(b-c)^2} \ln \frac{c}{b} 
\, ,
\label{eq:fc2}
\\
G^{c_3} (a,b,c,d) 
&\equiv &
\frac{2a^2 (a^2-bc) - a^2 (a-b)(a-c) }{2(a-b)^3(a-c)^3} \ln a \ln d 
+\frac{a d}{2(a-b)^2(a-c)^2} \left[ \ln \left( \frac{a}{d} \right) -1 \right]
\nonumber\\
&&
+\frac{ad (a^2 -bc)}{(a-b)^3(a-c)^3} \ln a 
-\frac{2(d-a)^2 (a^2 -bc) +(d^2-a^2)(a-b)(a-c)}{2(a-b)^3(a-c)^3} {\rm Li}_2 \left( 1-\frac{a}{d} \right) 
\nonumber\\
&&
+(\mbox{even permutations of } a, b, c)
\, ,
\label{eq:fc3}
\\
G^{c_4} (a,b,c,d) 
&\equiv &
\frac{(a -d)(a^2-bc-2ad+bd+cd)}{(a-b)^3(a-c)^3} {\rm Li}_2 \left( 1-\frac{a}{d} \right) 
-\frac{a (a^2-bc)}{(a-b)^3(a-c)^3} \ln a \ln d
+\frac{d (1+\ln d)}{2(a-b)^2(a-c)^2}
\nonumber\\
&&
+\frac{a d ( -2a+b+c)}{(a-b)^3(a-c)^3} \ln a
\nonumber\\
&&
+(\mbox{even permutations of } a, b, c)
\, .
\label{eq:fc4}
\end{eqnarray}
The derivation of this formula is given in Appendix \ref{sec:type1charginorainbow}.

Finally, the RPV rainbow contribution to the chromo-EDM of a down-type quark $D$ is given by
\begin{eqnarray}
d^{c ; {\rm rb}}_{D_k}
&=&
{\rm Im} ( \lambda'_{ijj} \lambda'^*_{ikk} e^{-i\delta_{d_j}} )
\frac{ \alpha_s }{256 \pi^3} m_{\tilde g} \sin \theta_{d_j} \cos \theta_{d_j} 
\sum_{\tilde D= \tilde D_L , \tilde D_R} 
G'(m_{\tilde g}^2, 0 , m_{\tilde D_{k}}^2 , m_{\tilde d_{1j}}^2 \, )
-(m_{\tilde d_{1j}}^2 \leftrightarrow m_{\tilde d_{2j}}^2)
\nonumber\\
&&
-
\sum_{n=1,2} 
\sum_{\tilde D= \tilde D_L , \tilde D_R} 
{\rm Im} ( \hat \lambda_{ijj} \lambda'^*_{ikk} e^{i(\theta_n -\delta_{f_j})})
\frac{n_c }{256 \pi^4} m_{\lambda_n } \sin \theta_{f_j} \cos \theta_{f_j} 
g^{(n)}_{\tilde D} 
\Bigl( g^{(n)}_{\tilde f_L} +g^{(n)}_{\tilde f_R} \Bigr)
m_{\tilde D_{k}}^2
G''(m_{\lambda_n}^2, 0 , m_{\tilde D_{k}}^2 , m_{\tilde f_{1j}}^2 \, )
\nonumber\\
&&
-(m_{\tilde f_{1j}}^2 \leftrightarrow m_{\tilde f_{2j}}^2)
\, ,
\label{eq:cedm}
\end{eqnarray}
\end{widetext}
where the inner loop fermion-sfermion pair ($f$ and $\tilde f$) of the second term may either be a down-type quark-squark loop or a charged lepton-slepton loop in RPV.
Here it is important to note that the inner lepton-slepton loop contribution has no analogue of Barr-Zee type diagrams [see Eq. (\ref{eq:chromo-Barr-Zee})], and it is thus a new characteristic process for the rainbow diagrams.
The phase of $m_{\tilde g}$ has already been fixed so as to make it real and positive.
We see that the formulae (\ref{eq:edmnug}) and (\ref{eq:cedm}) are very similar to the supersymmetric rainbow contribution with $R$-parity conservation \cite{Yamanaka:2012ia} (this is just a replacement of the matter-Higgs Yukawa couplings by the RPV couplings, and the Higgsino mass by the negligible mass of the neutrino).
The RPV rainbow diagrams vanish if the mass eigenvalues of the inner loop sfermions $m_{\tilde f_{1j}}$ and $m_{\tilde f_{2j}}$ are degenerate.
This property could also be seen for the supersymmetric rainbow process with $R$-parity conservation.

\section{\label{sec:many-body}From elementary level to observables}

The elementary level processes calculated in the previous section generates CP-odd operators renormalized at the scale of supersymmetric particles $\mu_{\rm SUSY} \sim O({\rm TeV})$.
Since most of the CP-odd hadron matrix elements are given at the renormalization point $\mu_{\rm had} =1$ GeV, we have to calculate the evolution down to this scale.
After the running of the energy scale, QCD operators change their Wilson coefficients in general.
While running the scale from 1 TeV to 1 GeV, the quark EDM evolves as \cite{Yamanaka:2017mef,Degrassi:2005zd}
\begin{equation}
d_q (\mu_{\rm had} ) 
=
0.8 \, d_q ({\rm TeV} ) 
.
\end{equation}
The scale of supersymmetry breaking may be much higher than 1 TeV, but the change from the above value is not important thanks to the asymptotic freedom (e.g. the running from 10 TeV to 1 TeV changes the coefficient by less than 5\%).
For the case of the quark chromo-EDM, we have to consider the mixing between operators since it also generates the quark EDM after evolution.
By defining the electric and chromo-electric dipole operators as
\begin{eqnarray}
{\cal L}_{d}
&=&
- \sum_q^{N_q} \frac{i Q_q e d_q}{2} \bar q \sigma_{\mu \nu} F^{\mu \nu} \gamma_5 q
\nonumber\\
&&
- \sum_q^{N_q} \frac{i g_s d^c_q}{2} \bar q \sigma_{\mu \nu} G_a^{\mu \nu} t_a \gamma_5 q
\nonumber\\
&\equiv &
\sum_q^{N_q} Q_q e d_q O_q
+
\sum_q^{N_q} g_s d^c_q O_q^c
,
\label{eq:dipoleoperators}
\end{eqnarray}
the evolution down to the hadron scale reads
\begin{equation}
O_q^c ({\rm TeV} ) 
\approx
-0.8 \, O_q (\mu_{\rm had} ) 
+0.9 \, O_q^c (\mu_{\rm had} ) 
.
\label{eq:cedmrunning}
\end{equation}
Here again the running from 10 TeV to 1 TeV changes the above coefficients by less than 5\% \cite{Degrassi:2005zd}.

Now let us move to the conversion of quark level CP violation to the hadron level effective interactions.
The quark EDM only contributes to the nucleon EDM in the leading order of electromagnetic interaction.
From recent lattice QCD calculations \cite{Yamanaka:2018uud,Gupta:2018lvp,Horkel:2020hpi,Tsuji:2022ric,Bali:2023sdi}, we have
\begin{eqnarray}
d_n 
&\approx &
0.8 d_d (\mu_{\rm had} ) 
-0.2 d_u (\mu_{\rm had} ) 
\nonumber\\
&\approx &
0.63 d_d ({\rm TeV} ) 
-0.16 d_u ({\rm TeV} ) 
,
\\
d_p 
&\approx &
0.8 d_u (\mu_{\rm had} ) 
-0.2 d_d (\mu_{\rm had} ) 
\nonumber\\
&\approx &
0.63 d_u ({\rm TeV} ) 
-0.16 d_d ({\rm TeV} ) 
,
\end{eqnarray}
with about 10\% of uncertainty.
The last equalities for each nucleon take into account the running from TeV scale.
The coefficients are called nucleon tensor charges, and their values show that the quark EDM is not enhanced at the hadronic level \cite{Yamanaka:2013zoa,Wang:2018kto,Liu:2019wzj}.
Their extractions from experimental data pointed to smaller values \cite{DAlesio:2020vtw,Gamberg:2021lgx,Boussarie:2023izj}, but recent analysis shows consistency \cite{Cocuzza:2023oam,Cocuzza:2023vqs}.

The quark chromo-EDM contributes to the nucleon EDM and also to the CP-odd pion-nucleon interaction \cite{Haxton:1983dq,Towner:1994qe}
\begin{equation}
{\cal L}_{\pi NN}^{(1)}
=
\bar g_{\pi NN}^{(1)}
\pi_0 \bar NN
,
\end{equation}
in the leading order of chiral effective field theory.
The NLO analysis of the isovector CP-odd pion-nucleon interaction yields \cite{Yamanaka:2016umw,Pospelov:2001ys,Bsaisou:2014zwa,deVries:2016jox,Osamura:2022rak,Sahoo:2023vrn}
\begin{eqnarray}
\bar{g}_{\pi N N}^{(1)} (\tilde d_q)
&\approx &
\Biggl[
\sigma_{\pi N}
+\frac{5 g_A^2 m_\pi^3}{64 \pi f_\pi^2}
\Biggr]
\frac{m_0^2 \bigl[d^c_d (\mu_{\rm had} ) -d^c_u (\mu_{\rm had} ) \bigr]}{2f_\pi (m_u + m_d) }
\ 
\nonumber\\
&= &
(125 \pm 75)
\bigl[d^c_d ({\rm TeV} ) -d^c_u ({\rm TeV} ) \bigr]
 \, {\rm fm}^{-1}
,
\label{eq:pion-nucleon_cEDM}
\end{eqnarray}
where $m_\pi = 138$ MeV, $f_\pi = 93$ MeV, and $g_A=1.27$.
At the scale $\mu_{\rm had} = 1$ GeV, the quark masses are $m_u = 2.9$ MeV and $m_d=6.0$ MeV \cite{Yamanaka:2015ncb}.
The last equality takes into account the running of the chromo-EDM [see Eq. (\ref{eq:cedmrunning})].
Here we only considered contributions generated by the dynamics of pions, since it is particularly enhanced by the pion-nucleon sigma-term $\sigma_{\pi N} \equiv (m_u+m_d) \langle N | \bar uu + \bar dd| N \rangle /2$ \cite{Yamanaka:2014lva,Huang:2019crt,Wang:2022rjv}, and we neglected the bare term (counterterm) which is estimated to be small \cite{Seng:2018wwp,Courtoy:2022kca}.
We adopt the value $\sigma_{\pi N} = (45 \pm 15) $ MeV for which the systematic error is due to the difference between lattice results \cite{Yamanaka:2018uud,Gupta:2021ahb,Agadjanov:2023jha,Bali:2023sdi} and phenomenological extractions \cite{Huang:2019not,Hoferichter:2023ptl}.
The mixed condensate $m_0^2 \equiv \langle 0 | \bar q g_s \sigma_{\mu \nu} G^{\mu \nu}_a t_a q | 0 \rangle / \langle 0 | \bar q q | 0 \rangle = (0.8 \pm 0.2 )$ GeV$^2$ was determined from QCD sum rules \cite{Belyaev:1982sa,Ioffe:2005ym,Gubler:2018ctz}.
To be conservative, we attribute an overall errorbar of 60\%.
We do not consider the quark EDM generated by the renormalization of the chromo-EDM because it has a small contribution compared to the chromo-EDM effect which is enhanced at the hadronic level.

The isovector CP-odd pion-nucleon interaction is also generated by the CP-odd four-quark interaction (\ref{eq:4fint}).
Using the factorization approximation, we have \cite{Gudkov:1992yc}
\begin{eqnarray}
\bar{g}_{\pi N N}^{(1)} ( C_{sd})
&\approx &
-\frac{f_\pi m_\pi^2}{m_u+m_d}
\langle N|\bar ss |N \rangle C_{sd} (\mu_{\rm had} )
\nonumber\\
&= &
-0.35 \, {\rm GeV}^2 \,C_{sd} ({\rm TeV} )
.
\end{eqnarray}
The strange content of the nucleon is obtained from lattice QCD calculations \cite{Alexandrou:2019brg}, $\sigma_{s} \equiv m_s \langle N | \bar ss| N \rangle \approx 50 $ MeV ($m_s = 120$ MeV at $\mu_{\rm had} = 1$ GeV \cite{Yamanaka:2015ncb}).
Here we neglect the effect of the bottom quark because the contribution from the Barr-Zee type diagrams is much more important.

The strongest constraint on hadronic CP violation is currently given by the EDM of $^{199}$Hg atom ($|d_{\rm Hg}|  < 7.4 \times 10^{-30}e$ cm) \cite{Graner:2016ses}.
The nucleon EDM and the CP violating pion-nucleon interaction contribute to the nuclear Schiff moment \cite{Yanase:2020agg,Yanase:2020oos,Yanase:2023jdr}
\begin{equation}
S_{\rm Hg}
=
2.65 d_n \, {\rm fm}^2
- 0.075 g_{\pi NN} \bar g_{\pi NN}^{(1)} e\, {\rm fm}^3
,
\label{eq:schiffmoment}
\end{equation}
where $g_{\pi NN} = g_A m_N / f_\pi \approx 12.8$.
The nuclear level uncertainty is estimated to be around  30\%.
Finally, after the atomic level calculation, we obtain the EDM of $^{199}$Hg atom \cite{Sahoo:2018ile,Hubert:2022pnl}:
\begin{equation}
d_{\rm Hg}
=
-2.4 \times 10^{-17}
\frac{S_{\rm Hg}}{{\rm fm}^3} {\rm cm}
.
\end{equation}

Combining all hadronic contributions, we have
\begin{eqnarray}
d_{\rm Hg} 
&=& 
1.1 \times 10^{-4} d_u 
-4.2 \times 10^{-4} d_d 
\nonumber\\
&&
- 2.9 \times 10^{-2} e ( d^c_u- d^c_d)
\nonumber\\
&&
+8 \times 10^{-18} C_{sd} \, {\rm GeV}^2 e \, {\rm cm}
,
\label{eq:HgEDM}
\end{eqnarray}
where all Wilson coefficients are renormalized at $\mu = 1$ TeV.

In this analysis, the CP-odd electron-nucleon (e-N) interaction
\begin{equation}
{\cal L}_{eN}
=
-\frac{G_F}{\sqrt{2}} 
\sum_{N=p,n} 
C_N^{SP} \bar N N \, \bar e i \gamma_5 e
,
\label{eq:e-N}
\end{equation}
where $G_F$ is the Fermi constant, is also important.
The relation between $C_N^{SP}$ and the four-fermion coupling of Eq. (\ref{eq:4fint}) is 
\begin{equation}
C_N^{SP}
=
-\sum_{q=d,s,b} \frac{\sqrt{2}C_{qe}}{G_F} \langle N | \bar q q| N \rangle
,
\label{eq:CSP}
\end{equation}
where we again need the nucleon scalar matrix elements $\langle N | \bar q q| N \rangle$.
The down- and strange quark contributions were already given above using $\sigma_{\pi N}$ and $\sigma_s$.
Here the bottom content $\langle N | \bar bb | N \rangle \approx 0.014$ (at the renormalization scale $\mu = m_b$) is also relevant.

The molecular beam experiment measures the frequency shift due to the electron EDM or the CP-odd e-N interaction.
For the HfF$^+$ ion, we have \cite{Petrov:2007zz,Fleig:2017mls}
\begin{equation}
\Delta f_{\rm HfF^+}
= 
\Biggl(
3.5 \times 10^{25} \frac{d_e}{e\, {\rm cm}}
+3.2 \times 10^5 C_N^{SP}
\Biggr)
\, {\rm rad /s}
.
\end{equation}
The current experimental bound on the electron EDM is $|d_e|  < 4.1 \times 10^{-30}e$ cm \cite{Roussy:2022cmp}, and equivalently $|C_N^{SP}|  < 4.5 \times 10^{-10}$ under the assumption of the single coupling dominance.

\section{\label{sec:analysis}Analysis}

\subsection{\label{sec:rainbowvsbarrzee}Comparison of Rainbow and Barr-Zee type diagrams}

We now analyze the contribution of the RPV rainbow diagrams to the EDM and to the chromo-EDM of fermions.
We choose the following values for the sparticle masses in the function of the supersymmetry breaking scale $\Lambda_{\rm SUSY}$:
$m_{\tilde d_R} \approx m_{\tilde d_{1}} = 2.8 \times \Lambda_{\rm SUSY}$, $m_{\tilde u_L} \approx m_{\tilde d_L} \approx m_{\tilde d_{2}} = 2.9 \times \Lambda_{\rm SUSY}$, $m_{\tilde g} = 3.0\times \Lambda_{\rm SUSY}$, $m_{\tilde e_R} \approx m_{\tilde e_{1}} = 0.9 \times \Lambda_{\rm SUSY}$, and $m_{\tilde \nu} \approx m_{\tilde e_L} \approx m_{\tilde e_{2}} = 1.0 \times \Lambda_{\rm SUSY}$.
We assume that sparticle masses are generation independent.
The $U(1)_Y$ and $SU(2)_L$ gaugino masses are given by $m_{\lambda_{1,2}} = 1.0 \times \Lambda_{\rm SUSY}$.
In our discussion, we fix  the $\mu$-parameter to $\mu = 1$ TeV.
For $\tan \beta$, we assign $\tan \beta =40$.
The large value of $\tan \beta$ is important, since the rainbow diagrams will be enhanced for large $\tan \beta$, due to the factor of $\cos \theta_{f}$.
The rainbow contribution will be compared with the RPV Barr-Zee type diagram (\ref{eq:BZtot}) [or with Eq. (\ref{eq:chromo-Barr-Zee}) for the chromo-EDM].

\begin{figure}[htb]
\begin{center}
\includegraphics[width=8cm]{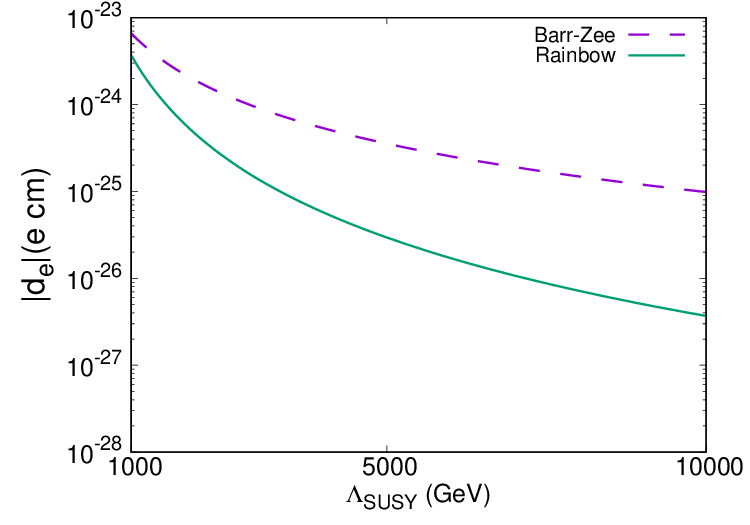}
\caption{\label{fig:rpv_l-lrbeedm}
Contribution of the RPV rainbowlike diagram (absolute value) to the electron EDM with third generation lepton-slepton inner loop plotted in the function of the supersymmetry breaking scale.
The Barr-Zee type contribution (absolute value) is also plotted for comparison.
We have set ${\rm Im}(\lambda_{233} \lambda^*_{211} e^{i(\theta_n -\delta_\tau)})= {\rm Im}(\lambda_{233} \lambda^*_{211})=1$.
}
\end{center}
\end{figure}

\begin{figure}[htb]
\begin{center}
\includegraphics[width=8cm]{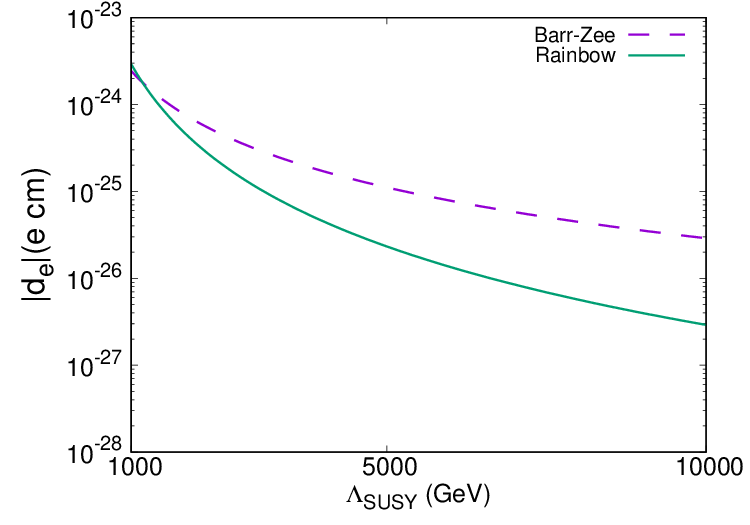}
\caption{\label{fig:rpv_l-qrbeedm}
Contribution of the RPV rainbowlike diagram (absolute value) to the electron EDM with third generation quark-squark inner loop to the electron EDM plotted in the function of the supersymmetry breaking scale.
The Barr-Zee type contribution (absolute value) is also plotted for comparison.
We have set ${\rm Im}(\lambda'_{i33} \lambda^*_{i11} e^{i(\theta_n -\delta_b)}) = {\rm Im}(\lambda'_{i33} \lambda^*_{i11})=1$.
}
\end{center}
\end{figure}

\begin{figure}[htb]
\begin{center}
\includegraphics[width=8cm]{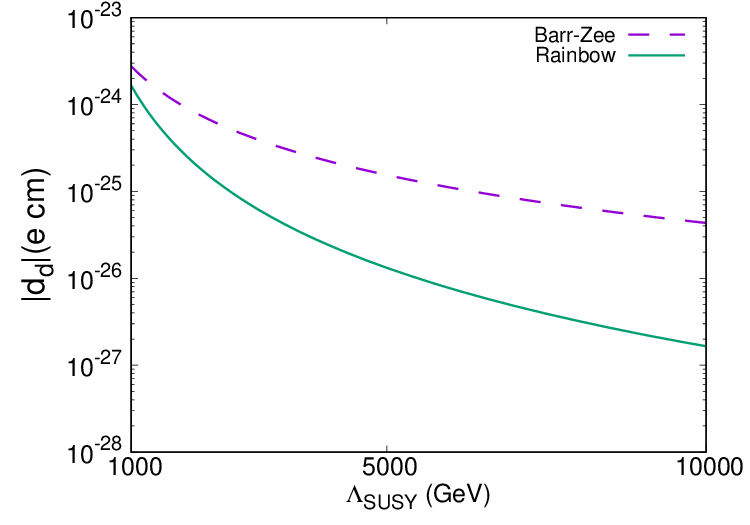}
\caption{\label{fig:rpv_q-lrbdedm}
Contribution of the RPV rainbow diagram (absolute value) to the down-quark EDM with third generation lepton-slepton inner loop plotted in the function of the supersymmetry breaking scale.
The Barr-Zee type contribution (absolute value) is also plotted for comparison.
We have set ${\rm Im}(\lambda_{i33} \lambda'^*_{i11} e^{i(\theta_n -\delta_\tau)}) = {\rm Im}(\lambda_{i33} \lambda'^*_{i11})=1$.
}
\end{center}
\end{figure}

\begin{figure}[htb]
\begin{center}
\includegraphics[width=8cm]{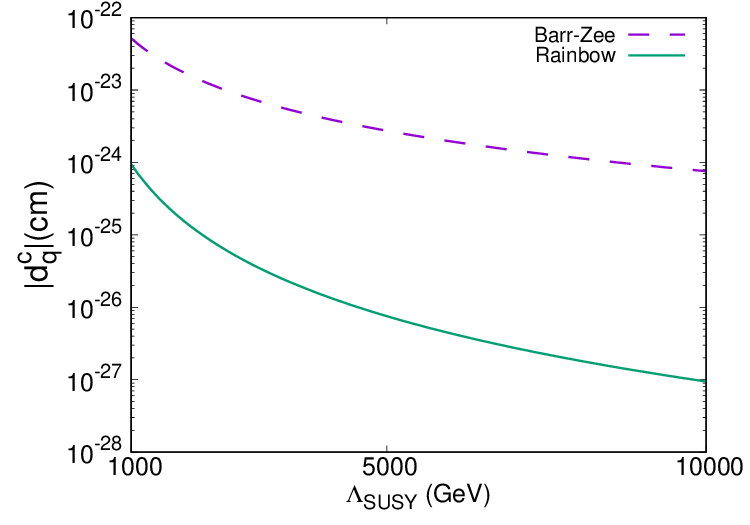}
\caption{\label{fig:rpv_q-qrbcedm}
Contribution of the RPV rainbowlike diagram (absolute value) to the chromo-EDM of down-quark with third generation quark-squark inner loop plotted in the function of the supersymmetry breaking scale.
The Barr-Zee type contribution (absolute value) is also plotted for comparison.
We have set ${\rm Im}(\lambda'_{i33} \lambda'^*_{i11} e^{i(\theta_n -\delta_b)}) =  {\rm Im}(\lambda'_{i33} \lambda'^*_{i11} e^{ -i\delta_b}) = {\rm Im}(\lambda'_{i33} \lambda'^*_{i11})=1$.
}
\end{center}
\end{figure}

In Figs. \ref{fig:rpv_l-lrbeedm}, \ref{fig:rpv_l-qrbeedm}, \ref{fig:rpv_q-lrbdedm}, and \ref{fig:rpv_q-qrbcedm}, we plot the $\Lambda_{\rm SUSY}$-dependence of the rainbow diagram contribution to the electron EDM generated by $\lambda_{233} \lambda^*_{211}$, $\lambda'_{i33} \lambda^*_{i11}$ ($i=2,3$), the down-quark EDM by $\lambda_{i33} \lambda'^*_{i11}$ ($i=1,2$), and the chromo-EDM of down-quark due to $\lambda'_{i33} \lambda'^*_{i11}$ ($i=1,2,3$), respectively, together with the Barr-Zee type one.
Regarding $\lambda'_{i33} \lambda'^*_{i11}$, we focused on the chromo-EDM since this is the most important process through which the EDM of $^{199}$Hg sets the strongest constraint on RPV couplings.
As a general observation, we see that the rainbowlike diagrams are smaller than the Barr-Zee type ones.
However, at the supersymmetry breaking scale $\Lambda_{\rm SUSY} = 1$ TeV, both effects are comparable, and the rainbow contribution even represents the largest effect as regards $\lambda'_{i33} \lambda^*_{i11}$ (see Fig. \ref{fig:rpv_l-qrbeedm}).
Here we note that the rainbow contribution is composed of $d_{F_k}^{\chi_0}$ and $d_{F_k}^{\chi_-}$ having opposite sign, so there are other parameter setups which generate an EDM larger than $d_{F_k}^{\rm BZ}$ in a wider region of $\Lambda_{\rm SUSY}$.
The rainbow diagrams drop faster than the Barr-Zee type ones as we increase $\Lambda_{\rm SUSY}$.
From Eqs. (\ref{eq:Barr-Zee}), (\ref{eq:ZBZ}), (\ref{eq:WBZ}), and (\ref{eq:chromo-Barr-Zee}), we see that the RPV Barr-Zee type diagrams scale as $\Lambda_{\rm SUSY}^{-2}$.
The RPV rainbow diagrams however scale as $\Lambda_{\rm SUSY}^{-3}$, due to $\cos \theta_f \propto \frac{m_f  \mu}{m_{\tilde f_L}^2 - m_{\tilde f_R}^2}$, where we have fixed the $\mu$-parameter.
If $\mu$ has a supersymmetric dynamical origin \cite{Giudice:1988yz} so that it scales as $\Lambda_{\rm SUSY}$, the rainbow diagram contribution will scale as $\Lambda_{\rm SUSY}^{-2}$ like the Barr-Zee type diagram, and it might be possible that they will stay comparable at high $\Lambda_{\rm SUSY}$.
Another remarkable point is that in RPV, the rainbow contribution is enhanced by tan $\beta$ while the Barr-Zee type diagrams are not.
The relative magnitude between them will therefore strongly depend on tan $\beta$.
We must finally note that the rainbow diagrams depend on additional CP phases $e^{i(\theta_n -\delta_f)}$, $e^{-i\delta_f}$, so we cannot know in advance whether they will interfere constructively or destructively with the Barr-Zee type diagrams.
Therefore, to set conservative upper limits on the combination of RPV couplings, we have to consider the destructive interference.

\subsection{\label{sec:RPVconstraint}Constraints on RPV couplings}

We now discuss the constraints on RPV couplings.
The combinations $\lambda_{322} \lambda^*_{311}$ and $\lambda_{233} \lambda^*_{211}$ (called {\it type 1} in Refs. \cite{Yamanaka:2014mda,Yamanaka:2014nba}) can only be probed by the electron EDM, so the experimental data of HfF$^+$ ion $|d_e| < 4.1 \times 10^{-30} e\, {\rm cm}$ \cite{Roussy:2022cmp} directly constrain them.
The final upper limits are shown in Table \ref{table:RPVlimits}.

The second class of RPV couplings to be inspected is the {\it type 2}, composed of $\lambda_{i11} \lambda^*_{i11}$, $\lambda_{i22} \lambda^*_{i11}$ and $\lambda_{i33} \lambda^*_{i11}$ ($i=2,3$).
These also contribute to the electron EDM, but it is actually known that the CP-odd e-N interaction [Eq. (\ref{eq:e-N})] has a larger effect on the frequency shift measured in HfF$^+$ ion experiment.
The latter has a larger contribution even by taking into account the theoretical uncertainty of 40\% in the evaluation of the hadron level coefficients (\ref{eq:CSP}).
The upper limits on RPV couplings are thus given by the CP-odd e-N interaction, as shown in Table \ref{table:RPVlimits}.

The third case, {\it type 3}, involving the RPV couplings $\lambda_{j22} \lambda'^*_{j11}$ ($j=1,3$) and $\lambda_{k33} \lambda'^*_{k11}$ ($k=1,2$), needs detailed discussion.
This class of RPV couplings actually contributes to the down-quark EDM, and also to the chromo-EDM via the second term of Eq. (\ref{eq:cedm}).
The latter is characteristic of the rainbow diagram, so no interference with the Barr-Zee type contribution exists.
The effect of the chromo-EDM is enhanced compared to the quark EDM, so the type 3 RPV couplings are likely to be constrained by the experimental data of $^{199}$Hg EDM ($|d_{\rm Hg}| < 7.4 \times 10^{-30} e\, {\rm cm}$ \cite{Graner:2016ses}) via this process.
With $\Lambda_{\rm SUSY} = 1$ TeV, this is indeed the case (see Table \ref{table:RPVlimits}).
For higher $\Lambda_{\rm SUSY}$, however, the effect of rainbow diagrams drops as explained in Sec. \ref{sec:rainbowvsbarrzee}, and the down-quark EDM, which involves Barr-Zee type diagrams, becomes the leading contribution.
There the type 3 RPV couplings are constrained by the neutron EDM ($|d_n| < 1.8 \times 10^{-26} e\, {\rm cm}$ \cite{Abel:2020gbr}) and by the $^{199}$Hg EDM experiments.
It happens that the two experiments are currently giving very close upper limits, with the $^{199}$Hg EDM providing a slightly stronger one, if we assume theoretical uncertainties of 10\% for the neutron and 40\% for $^{199}$Hg (see Section \ref{sec:many-body}).
The upper limits given by the $^{199}$Hg EDM experiment are shown in Table \ref{table:RPVlimits}.
We see that the constraints are not very strong.
Regarding the combination $\lambda_{122} \lambda'^*_{111}$, there are actually tighter limits provided by other processes.
From the analysis of the universality of charged lepton decay, we have \cite{Barbier:2004ez,Kao:2009fg}
\begin{eqnarray}
|\lambda_{122} | & < & 0.3 \times \frac{m_{\tilde e_{R}}}{1 \, {\rm TeV}}
,
\end{eqnarray}
while we also have a strong constraint from the neutrinoless double beta decay \cite{Faessler:2007nz}
\begin{eqnarray}
|\lambda'_{111} | & < & 8.9 \times 10^{-2} \times \left( \frac{m_{\tilde d}}{1 \, {\rm TeV}} \right)^{\frac{3}{2}}
.
\end{eqnarray}
As we can see, these constraints yield lower upper limits on RPV couplings than the EDM.

The final class is the {\it type 4} RPV, $\lambda'_{l22} \lambda'^*_{l11}$ and $\lambda'_{l33} \lambda'^*_{l11}$ ($l=1,2,3$) which is probed by the $^{199}$Hg EDM.
The couplings $\lambda'_{l22} \lambda'^*_{l11}$ contribute via the CP-odd 4-quark interaction, and $\lambda'_{l33} \lambda'^*_{l11}$ through the quark chromo-EDM.
They also generate the quark EDM, but the effect is much smaller than the above purely hadronic processes.
We see from the result displayed in Table \ref{table:RPVlimits} that, although affected by large errorbars due to nonperturbative effects of QCD, the upper limits provided by the $^{199}$Hg EDM experimental data is quite strong.

\begin{table}
\caption{
Upper limits on RPV couplings set from this analysis for two selections of the supersymmetry breaking scale $\Lambda_{\rm SUSY}$.
We assumed that the rainbow and Barr-Zee type diagrams add with opposite sign.
In this Table, $i=2,3$, $j=1,3$, $k=1,2$, and $l=1,2,3$.
The symbol ``$-$'' means that the upper limit exceeds one and that we could not set any constraints.
}
\begin{ruledtabular}
\begin{tabular}{lcc}
RPV couplings & $\Lambda_{\rm SUSY}=1$ TeV & $\Lambda_{\rm SUSY}=10$ TeV \\
\hline
$| {\rm Im}(\lambda_{322} \lambda^*_{311}) |$ & $1.2 \times 10^{-5}$ & $5.5 \times 10^{-4}$ \\
$| {\rm Im}(\lambda_{233} \lambda^*_{211}) |$ & $1.4 \times 10^{-6}$ & $4.3 \times 10^{-5}$ \\
\hline
$| {\rm Im}(\lambda'_{i11} \lambda^*_{i11}) |$ & $5.0 \times 10^{-10}$ & $4.4 \times 10^{-8}$ \\
$| {\rm Im}(\lambda'_{i22} \lambda^*_{i11}) |$ & $9.1 \times 10^{-9}$ & $8.1 \times 10^{-7}$ \\
$| {\rm Im}(\lambda'_{i33} \lambda^*_{i11}) |$ & $2.3 \times 10^{-7}$ & $2.1 \times 10^{-5}$ \\
\hline
$| {\rm Im}(\lambda_{j22} \lambda'^*_{j11}) |$ & $3.8 \times 10^{-2}$ & $-$ \\
$| {\rm Im}(\lambda_{k33} \lambda'^*_{k11}) |$ & $2.2 \times 10^{-3}$ & $7.3 \times 10^{-1}$ \\
\hline
$| {\rm Im}(\lambda'_{l22} \lambda'^*_{l11}) |$ & $2.3 \times 10^{-6}$ & $1.8 \times 10^{-4}$ \\
$| {\rm Im}(\lambda'_{l33} \lambda'^*_{l11}) |$ & $1.4 \times 10^{-5}$ & $9.8 \times 10^{-4}$ \\
\end{tabular}
\end{ruledtabular}

\label{table:RPVlimits}
\end{table}

\section{Conclusion}

In this paper, we calculated new sizable RPV contributions to the fermion EDM, the rainbow diagrams.
The calculation was similar to that of the minimal supersymmetric standard model with R-parity conservation, and we have given explicit formulae in this paper.
It was found that the RPV rainbow diagrams are smaller than the RPV Barr-Zee type ones in most of the cases, but comparable at the supersymmetry breaking scale close to 1 TeV.
They also have many interesting properties, such as the large enhancement due to $\tan \beta$, the $\Lambda_{\rm SUSY}^{-3}$ scaling, etc.
The interference between the Barr-Zee type and rainbow diagrams is dependent on the phases $e^{i(\theta_n -\delta_{f_j})}$, $e^{-i\delta_{q_j}}$ and the signs of $\cos \theta_{f_j} \sin \theta_{f_j}$.
To analyze the constraints on RPV couplings that can be imposed from EDM experiments, we assumed that the two effects are maximally destructive, and we could set conservative upper limits.
We note that if the $\mu$-parameter is generated by the dynamics of supersymmetry breaking, the scaling of rainbow diagrams may become $\Lambda_{\rm SUSY}^{-2}$ so that they will stay comparable with the Barr-Zee type diagrams for any $\Lambda_{\rm SUSY}$, but this is only a naive estimation, and this needs an explicit investigation to be confirmed.

The combinations of RPV couplings that were analyzed in this work can be classified into four categories.
The type 1 ($\lambda_{322} \lambda^*_{311}$ and $\lambda_{233} \lambda^*_{211}$) and type 2 [$\lambda_{i11} \lambda^*_{i11}$, $\lambda_{i22} \lambda^*_{i11}$ and $\lambda_{i33} \lambda^*_{i11}$ ($i=2,3$)] are constrained by the HfF$^+$ ion experimental data, each via the electron EDM and the CP-odd e-N interaction.
They may further be constrained by improving the sensitivity of paramagnetic molecular experiments, with several new ideas \cite{Sakemi:2011zz,Vutha:2018tsz,Zakharova:2020fam,Chekhovskoi:2022soa,Chamorro:2022hfb,Maison:2022miu}.

Regarding the type 3 RPV couplings [$\lambda_{j22} \lambda'^*_{j11}$ ($j=1,3$) and $\lambda_{k33} \lambda'^*_{k11}$ ($k=1,2$)], we found that the rainbow diagrams yield the largest contribution to the EDM of $^{199}$Hg atom via the chromo-EDM for $\Lambda_{\rm SUSY}=1$ TeV thanks to the absence of the Barr-Zee type analogue, and we could eventually set the strongest upper limits to $\lambda_{322} \lambda'^*_{311}$ and to $\lambda_{k33} \lambda'^*_{k11}$ ($k=1,2$), while $\lambda_{122} \lambda'^*_{111}$ is bounded by other experiments.
For high scale supersymmetry breaking, the down-quark EDM becomes important.
The improvement of the sensitivity of the neutron EDM experiment will therefore further constrain the type 3 RPV couplings.
Another attractive approach is to measure the EDM of the proton using storage rings \cite{Anastassopoulos:2015ura,bnl,Porshnev:2021xdz,Suleiman:2021whz} for which the prospect is to reach the sensitivity of $O(10^{-29})e$ cm.

The purely hadronic RPV, type 4, [$| {\rm Im}(\lambda'_{l22} \lambda'^*_{l11}) |$ and $| {\rm Im}(\lambda'_{l33} \lambda'^*_{l11}) |$ ($k=1,2,3$)] is strongly constrained by the $^{199}$Hg EDM experiment thanks to the enhancement of the chromo-EDM and CP-odd 4-quark interaction at the hadronic and nuclear levels.
To further constrain the type 4 RPV, nuclear systems are therefore well suited.
In this regard, there are also many on-going projects and new ideas of future experiments \cite{Bowman:2014fca,Yamanaka:2015qfa,Gudkov:2017vqn,Yamanaka:2019vec,Gudkov:2019ltu,Skripnikov:2020qgt,Flambaum:2020xcj,Froese:2021civ,Terrano:2021zyh,Snow:2021uuo,Scheck:2022bvc,Dalton:2023kfz}.

The main importance of this paper is the quantification of the RPV CP violation.
Since the EDM is a baryon and lepton number conserving quantity, it impossible to discriminate the RPV and R-parity conserving sectors using a single EDM experiment.
Fortunately, the dependences of EDMs of various systems on CP phases of R-parity conservation and violation are different, so it is in principle possible to separate their sources of CP violation by knowing all coefficients relating the elementary level CP phases to observable EDMs, and measuring the EDMs of an equal amount of different systems as the number of unknown CP phases.
The calculation and the quantification of elementary level diagrams were needed for this reason.
We also note that constraints on RPV couplings obtained in previous works \cite{Yamanaka:2012ep,Yamanaka:2014nba} were not accurate partly because the hadron level CP-odd low energy constants and nuclear level coefficients were almost unknown until recently.
The improvement of the analysis of hadronic RPV couplings was mainly due to the quantifications of the chromo-EDM [Eq. (\ref{eq:pion-nucleon_cEDM})] and the nuclear Schiff moment [Eq. (\ref{eq:schiffmoment})] \cite{Yanase:2020agg,Yanase:2020oos,Yanase:2023jdr}.

\begin{acknowledgments}
The author thanks H. Kamano for useful discussion and comments.
This work was supported by Daiko Foundation.
\end{acknowledgments}

\appendix
\onecolumngrid

\section{\label{sec:type1rainbow}Calculation of rainbow diagrams with a neutralino or a gluino with an effective gauge boson-neutrino-gaugino vertex}

We give the detail of the calculation of the rainbow diagrams with the insertion of the one-loop effective neutrino-gaugino-gauge boson vertex [Fig. \ref{fig:rainbow_type} (a)].
This contribution can generate both the EDM and the chromo-EDM.
\begin{figure}[htb]
\includegraphics[width=18cm]{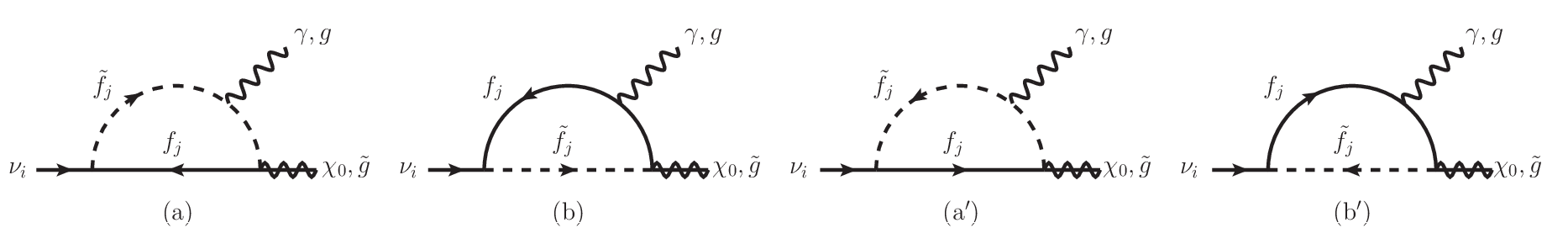}
\caption{\label{fig:nugg} 
One-loop level contribution to the effective neutrino-gaugino-gauge boson vertex.
}
\end{figure}
The amplitude of the effective neutrino-gaugino-gauge boson vertex is the sum of diagrams shown in Fig. \ref{fig:nugg}:
\begin{equation}
{\cal M}_{\nu \chi G} = 
{\cal M}_{\nu \chi G}^{\rm (a)} 
+ {\cal M}_{\nu \chi G}^{\rm (b)} 
+ {\cal M}_{\nu \chi G}^{\rm (a')} 
+ {\cal M}_{\nu \chi G}^{\rm (b')} \ .
\end{equation}
The explicit expressions of the off-shell amplitudes ${\cal M}_{\nu \chi G}^{\rm (a)}$ and ${\cal M}_{\nu \chi G}^{\rm (b)}$ are:
\begin{eqnarray}
i{\cal M}_{\nu \chi G}^{\rm (a)}
&=&
\hat \lambda_{ijj} \sqrt{2} g_f t_n \epsilon_\mu^* (q_1) \int \frac{d^4 k}{(2\pi )^4} \nonumber\\
&&\times 
(-2)g^{(n)}_{\tilde f_L} \bar \lambda_n (k\hspace{-.5em}/\, +q\hspace{-.5em}/_2 ) P_L \nu_i \cdot k^\mu \sin \theta_{f_j} \cos \theta_{f_j} e^{i\frac{\theta_n}{2} - i\delta_{f_j}}
\left[ d( m_{\tilde f_{2j}}^2 , m_{\tilde f_{2j}}^2 , m_{ f_j}^2 )
- d( m_{\tilde f_{1j}}^2 , m_{\tilde f_{1j}}^2 , m_{ f_j}^2 ) \right]
\nonumber\\
&\approx&
-\sqrt{2} t_n \hat \lambda_{ijj} g_f g^{(n)}_{\tilde f_L} \epsilon_\mu^* (q_1) 
\sin \theta_{f_j} \cos \theta_{f_j} e^{i\frac{\theta_n}{2} - i\delta_{f_j}}
\frac{i}{(4\pi)^2}
\nonumber\\
&&\times \int_0^1 dz\,\bar \lambda_n \Biggl[
-\gamma^\mu (1-z)\ln \left( z q_2^2 - m_{\tilde f_{2i}}^2 \right) 
- \frac{z(1-z)(q_1\cdot q_2 ) \gamma^\mu }{zq_2^2 - m_{\tilde f_{2i}}^2 }
\nonumber\\
&& \hspace{6em}
+
\frac{z (1-z) q_2^\mu ( q\hspace{-.5em}/_1 + 2 q\hspace{-.5em}/_2 )}{ -z q_2^2 + m_{\tilde f_{2i}}^2 } 
+2\frac{ z^2(1-z) (q_1\cdot q_2 )q_2^\mu q\hspace{-.5em}/_2 }{[ z q_2^2 - m_{\tilde f_{2i}}^2 ]^2}
\Biggr] P_L \nu_i
-( m_{\tilde f_{2i}}^2 \leftrightarrow m_{\tilde f_{1i}}^2)  
\, ,
\label{eq:m(a)}
\end{eqnarray}
\begin{eqnarray}
i{\cal M}_{\nu \chi G}^{\rm (b)}
&=&
\hat \lambda_{ijj} \sqrt{2} g_f t_n \epsilon_\mu^* (q_1) \int \frac{d^4 k}{(2\pi )^4} 
\nonumber\\
&&\times 
(- g^{(n)}_{\tilde f_L} ) \bar \lambda_n (k\hspace{-.5em}/\, \gamma^\mu (k\hspace{-.5em}/\, - q\hspace{-.5em}/_1) + m_{f_j}^2 \gamma^\mu) P_L \nu_i \cos \theta_{f_j} \sin \theta_{f_j}  e^{i\frac{\theta_n}{2} - i\delta_{f_j}}
\left[ d( m_{ f_j}^2, m_{ f_j}^2 , m_{\tilde f_{2j}}^2 )
- d( m_{ f_j}^2, m_{ f_j}^2, m_{\tilde f_{1j}}^2  ) \right] 
\nonumber\\
&\approx &
- \sqrt{2} \hat \lambda_{ijj} t_n g_f g^{(n)}_{\tilde f_L} \cos \theta_{f_j} \sin \theta_{f_j}  e^{i\frac{\theta_n}{2} - i\delta_{f_j}}
\epsilon_\mu^* (q_1) \frac{i}{(4\pi)^2}
\nonumber\\
&&\times 
\int_0^1 dz \, \bar \lambda_n \Biggl[
\gamma^\mu z \ln \left( zq_2^2 - m_{\tilde f_{2i}}^2 \right)
+
\frac{
-z q\hspace{-.5em}/_2 q\hspace{-.5em}/_1 \gamma^\mu 
+ z^2 [ 2(q_1 \cdot q_2)\gamma^\mu -q_2^\mu q\hspace{-.5em}/_1 ]
+ z (1-z)(2q_2^\mu q\hspace{-.5em}/_2 -q_2^2 \gamma^\mu )
}{z q_2^2 - m_{\tilde f_{2i}}^2 }
\nonumber\\
&&\hspace{10em}
-\frac{
z^2 (1-z)(q_1\cdot q_2 )
(2q_2^\mu q\hspace{-.5em}/_2 -q_2^2 \gamma^\mu )
}{[ z q_2^2 - m_{\tilde f_{2i}}^2 ]^2}
\Biggr] P_L \nu_i 
-( m_{\tilde f_{2i}}^2 \leftrightarrow m_{\tilde f_{1i}}^2) 
,
\label{eq:m(b)}
\end{eqnarray}
where $g_f =Q_f e$ is the electromagnetic gauge coupling of the fermion $f$ or of the sfermion $\tilde f$ of the inner loop (the QCD coupling $g_f = g_s$ for the chromo-EDM), and $g_{\tilde f_{L/R}}$ the gaugino-fermion-sfermion coupling [see Eq. (\ref{eq:gaugino-int}) for the Lagrangian].
Here $\hat \lambda $ is the RPV coupling, with $\hat \lambda = \lambda $ ($= \lambda'$) when a charged lepton-slepton (down-type quark-squark) pair runs in the loop.
The factor $t_n$ is defined by $t_n = Q_f n_c$ for the contribution with an external photon (EDM) and $t_n = \frac{1}{2}$ with an external gluon (chromo-EDM).
The momenta carried by the exiting gauge boson and gaugino are denoted by $q_1$ and $q_2$, respectively.
The gaugino and the neutrino spinors are given by $\lambda_n$ and $ \nu_i$, respectively.
The polarization vector of the external gauge boson is $\epsilon^*_\mu$.
We have used the following notation to simplify the above equations:
\begin{equation}
d(x , y , z) \equiv 
 \frac{1}{\left[ k^2 - x \right] \left[ (k-q_1)^2 - y \right] \left[ (k+q_2)^2 - z \right]} \ .
\end{equation}
In Eqs. (\ref{eq:m(a)}) and (\ref{eq:m(b)}), we omitted terms proportional to $\cos^2 \theta_{f_j}$ and $\sin^2 \theta_{f_j}$ since they do not contribute to the EDM.
We have also neglected the quark and lepton masses.
The sum of Eqs. (\ref{eq:m(a)}) and (\ref{eq:m(b)}) becomes
\begin{eqnarray}
i{\cal M}_{\nu \chi G}^{\rm (a)}+i{\cal M}_{\nu \chi G}^{\rm (b)}
&\approx &
\frac{i\sqrt{2}}{(4\pi)^2} \hat \lambda_{ijj} t_n g_f g^{(n)}_{\tilde f_L} \sin \theta_{f_j} \cos \theta_{f_j} 
e^{i\frac{\theta_n}{2} - i\delta_{f_j}} \epsilon_\mu^* (q_1) 
\bar \lambda_n q\hspace{-.5em}/_1 q\hspace{-.5em}/_2 \gamma^\mu P_L \nu_i 
\int_0^1 dz \,\frac{z }{z q_2^2 - m_{\tilde f_{1j}}^2}
\nonumber\\
&&
-( m_{\tilde f_{1j}}^2 \leftrightarrow m_{\tilde f_{2j}}^2)
\, ,
\label{eq:m(a)+m(b)}
\end{eqnarray}
where we have further omitted several Dirac bilinears with Lorentz structures such as $q_2^\mu \bar \lambda_n q\hspace{-.5em}/_1 P_L \nu_i$ and $(q_1 \cdot q_2) \bar \lambda_n \gamma^\mu P_L \nu_i$ since these terms do not contribute to the EDM operator $\epsilon^*_\mu \bar F_k \sigma^{\mu \nu} q_\nu \gamma_5 F_k$ after insertion into the second loop (they cancel with their complex conjugates).
The logarithmic terms and that with a factor of ${z(1-z)q_2^2 }/{z q_2^2 - m_{\tilde f_{1,2}}^2}$ of Eqs. (\ref{eq:m(a)}) and (\ref{eq:m(b)}) cancel thanks to the following formula:
\begin{equation}
\int_0^1 \hspace{-.5em} dx\, \left[ ( 2x - 1) \ln \left( xa -b \right) - \frac{x(1-x)a}{ xa -b} \right]
=0
\, .
\label{eq:01integralformula}
\end{equation}
The amplitudes $i{\cal M}_{\nu \chi G}^{\rm (a')}$ and $i{\cal M}_{\nu \chi G}^{\rm (b')}$ can be calculated in a similar manner.
Summing them, we finally obtain the following one-loop level effective neutrino-gaugino-gauge boson vertex (expanded up to the first order in the external momentum carried by the gauge boson):
\begin{eqnarray}
i{\cal M}_{\nu \chi G}
&\approx &
\frac{i\sqrt{2}}{(4\pi)^2} \hat \lambda_{ijj} g_f t_n 
\Bigl(g^{(n)}_{\tilde f_L} -g^{(n)}_{\tilde f_R} \Bigr) 
\sin \theta_{f_j} \cos \theta_{f_j} 
e^{i\frac{\theta_n}{2} - i\delta_{f_j}} \epsilon_\mu^* (q_1) 
\bar \lambda_n q\hspace{-.5em}/_1 q\hspace{-.5em}/_2 \gamma^\mu P_L \nu_i 
\int_0^1 dz \,\frac{z }{z q_2^2 - m_{\tilde f_{1j}}^2}
\nonumber\\
&&
-( m_{\tilde f_{1j}}^2 \leftrightarrow m_{\tilde f_{2j}}^2)
\, ,
\label{eq:nulamG}
\end{eqnarray}
and the effective antineutrino-gaugino-gauge boson vertex
\begin{eqnarray}
i{\cal M}_{\nu^c \chi G}
&\approx &
\frac{i\sqrt{2}}{(4\pi)^2} \hat \lambda_{ijj}^* g_f t_n 
\Bigl(g^{(n)}_{\tilde f_R} -g^{(n)}_{\tilde f_L} \Bigr) 
\sin \theta_{f_j} \cos \theta_{f_j} e^{i\delta_{f_j}-i\frac{\theta_n}{2} } \epsilon_\mu^* (q_1) 
\bar \lambda_n q\hspace{-.5em}/_1 q\hspace{-.5em}/_2 \gamma^\mu P_R \nu_i^c 
\int_0^1 dz \,\frac{z }{z q_2^2 - m_{\tilde f_{1j}}^2}
\nonumber\\
&&
-( m_{\tilde f_{1j}}^2 \leftrightarrow m_{\tilde f_{2j}}^2)
\, .
\label{eq:antinulamG}
\end{eqnarray}
This amplitude is given by the complex conjugated RPV coupling $\hat \lambda_{ijj}^*$.
Note that we also have to consider the amplitudes generated by the transposed fermion interactions.
They are given by
\begin{eqnarray}
i{\cal M}_{\chi \nu G}
&\approx &
\frac{i\sqrt{2}}{(4\pi)^2} \hat \lambda^*_{ijj} g_f t_n 
\Bigl(g^{(n)}_{\tilde f_L} -g^{(n)}_{\tilde f_R} \Bigr) 
\sin \theta_{f_j} \cos \theta_{f_j} e^{i\delta_{f_j}-i\frac{\theta_n}{2} }\epsilon_\mu^* (q_1) 
\bar \nu_i q\hspace{-.5em}/_1 q\hspace{-.5em}/_2 \gamma^\mu P_L \lambda_n 
\int_0^1 dz \,\frac{z }{z q_2^2 - m_{\tilde f_{1j}}^2}
\nonumber\\
&&
-( m_{\tilde f_{1j}}^2 \leftrightarrow m_{\tilde f_{2j}}^2)
\, ,
\label{eq:lamnuG}
\\
i{\cal M}_{\chi \nu^c G}
&\approx &
\frac{i\sqrt{2}}{(4\pi)^2} \hat \lambda_{ijj} g_f t_n 
\Bigl(g^{(n)}_{\tilde f_R} -g^{(n)}_{\tilde f_L} \Bigr) 
\sin \theta_{f_j} \cos \theta_{f_j} e^{i\frac{\theta_n}{2} - i\delta_{f_j}} \epsilon_\mu^* (q_1) 
\bar \nu_i^c q\hspace{-.5em}/_1 q\hspace{-.5em}/_2 \gamma^\mu P_R \lambda_n 
\int_0^1 dz \,\frac{z }{z q_2^2 - m_{\tilde f_{1j}}^2}
\nonumber\\
&&
-( m_{\tilde f_{1j}}^2 \leftrightarrow m_{\tilde f_{2j}}^2)
\, .
\hspace{2em}
\label{eq:lamantinuG}
\end{eqnarray}
Here the momentum $q_2$ is carried by the exiting neutrino or antineutrino.
We should note that the one-loop level effective vertices (\ref{eq:nulamG}), (\ref{eq:antinulamG}), (\ref{eq:lamnuG}), and (\ref{eq:lamantinuG}) are all of first order in $q_1$.

We now insert the above effective vertices (\ref{eq:nulamG}), (\ref{eq:antinulamG}), (\ref{eq:lamnuG}), and (\ref{eq:lamantinuG}) into the second loop, as shown in Fig. \ref{fig:rainbow_type} (a).
We then obtain the following contribution (expanded up to the first order in $q_1$): 
\begin{eqnarray}
i{\cal M}^{\chi_0}_{\rm (a)}
&\approx &
\frac{i}{(4\pi)^2} {\rm Im} \Bigl[ \hat \lambda_{ijj} \tilde \lambda_{ikk}^* e^{i(\theta_n -\delta_{f_j})} \Bigr] g_f t_n \Bigl(g^{(n)}_{\tilde f_L} -g^{(n)}_{\tilde f_R} \Bigr) \sin \theta_{f_j} \cos \theta_{f_j} m_{\lambda_n} \epsilon_\mu^* (q_1) 
\nonumber\\
&& \times 
\sum_{\tilde F = \tilde F_L , \tilde F_R } s_{\tilde F} g_{\tilde F}^{(n)} \int \frac{d^4 k }{(2\pi)^4} 
\frac{-k^2 \cdot \bar F_k  q\hspace{-.5em}/_1 \gamma^\mu \gamma_5 F_k }{k^2 \left[ k^2 - m_{\lambda_n}^2 \right] \left[ k^2 - m_{\tilde F_k}^2 \right]}
\int_0^1 dz \,\frac{z }{z k^2 - m_{\tilde f_{1j}}^2}
-( m_{\tilde f_{1j}}^2 \leftrightarrow m_{\tilde f_{2j}}^2)
\nonumber\\
&\approx &
\frac{i}{256 \pi^4} {\rm Im} \Bigl[ \hat \lambda_{ijj} \tilde \lambda^*_{ikk} e^{i(\theta_n -\delta_{f_j})} \Bigr]
m_{\lambda_n } t_n g_f \Bigl(g^{(n)}_{\tilde f_R} -g^{(n)}_{\tilde f_L} \Bigr) \sin \theta_{f_j} \cos \theta_{f_j} 
\nonumber\\
&&\hspace{2em} \times
\epsilon_\mu^* (q_1) \bar F_k \sigma^{\mu \nu} (q_1)_\nu \gamma_5 F_k
\sum_{\tilde F= \tilde F_L , \tilde F_R} s_{\tilde F}\, g^{(n)}_{\tilde F} 
\int_0^\infty \hspace{-.5em}
\frac{r^2dr }{\Bigl[ r +m_{\lambda_n}^2 \Bigr] r  \Bigl[ r +m_{\tilde F_k}^2 \Bigr] } \int_0^1 \hspace{-0.5em} \frac{xdx}{xr + m_{\tilde f_{1j}}^2} 
-(m_{\tilde f_{1j}}^2 \leftrightarrow m_{\tilde f_{2j}}^2)
\nonumber\\
&= &
\frac{i}{256 \pi^4} {\rm Im} \Bigl[ \hat \lambda_{ijj} \tilde \lambda^*_{ikk} e^{i(\theta_n -\delta_{f_j})} \Bigr]
m_{\lambda_n } t_n g_f \Bigl(g^{(n)}_{\tilde f_R} -g^{(n)}_{\tilde f_L} \Bigr) \sin \theta_{f_j} \cos \theta_{f_j} 
\nonumber\\
&&\hspace{2em} \times
\epsilon_\mu^* (q_1) \bar F_k \sigma^{\mu \nu} (q_1)_\nu \gamma_5 F_k
\sum_{\tilde F= \tilde F_L , \tilde F_R} s_{\tilde F}\, g^{(n)}_{\tilde F} 
\left[ G'(m_{\lambda_n}^2, 0 , m_{\tilde F_{k}}^2 , m_{\tilde f_{1j}}^2 \, )
-G'(m_{\lambda_n}^2, 0 , m_{\tilde F_{k}}^2 , m_{\tilde f_{2j}}^2 \, ) \right]
, \ \ \ \ \ \ \ 
\end{eqnarray}
where we have omitted terms which do not contribute to the EDM operator $\epsilon^*_\mu \bar F_k \sigma^{\mu \nu} (q_1)_\nu \gamma_5 F_k$.
The RPV coupling $\tilde \lambda$ is $\tilde \lambda = \lambda$ ($= \lambda'$) for the EDM of charged lepton (down-type quark).
The constant $s_{\tilde F}$ is $+1$ for left-handed sfermion $\tilde F_L$ and $-1$ for right-handed sfermion $\tilde F_R$.
The function $G'$ is defined in Eqs. (\ref{eq:F'}).
From the above equation, we obtain the first term of the formula of the EDM (\ref{eq:edmnug}) and the chromo-EDM (\ref{eq:cedm}).

\section{\label{sec:type2rainbow}Calculation of rainbow diagrams with a neutralino or a gluino with an effective neutrino-gaugino transition}

\begin{figure}[htb]
\begin{center}
\includegraphics[width=9.2cm]{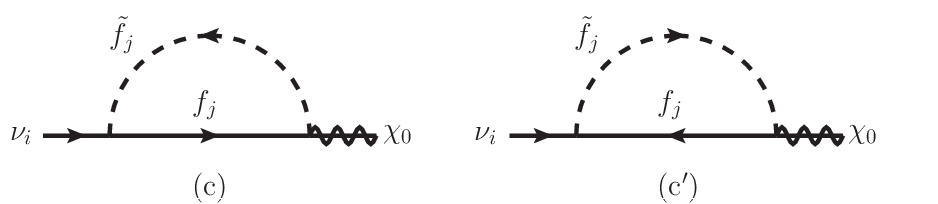}
\caption{\label{fig:nug}
One-loop level contribution to the RPV neutrino-gaugino transition.
}
\end{center}
\end{figure}

We calculate the rainbow diagrams with the insertion of the one-loop effective lepton-gaugino transition [Fig. \ref{fig:rainbow_type} (b)].
The contributing inner one-loop diagrams are shown in Fig. \ref{fig:nug}.
The one-loop level effective neutrino-gaugino transition within RPV is given by
\begin{eqnarray}
i{\cal M}_{\nu \chi}
&=&
- \sqrt{2} \hat \lambda_{ijj} n_c (g^{(n)}_{\tilde f_L} + g^{(n)}_{\tilde f_R }) \sin \theta_{f_j} \cos \theta_{f_j} e^{i\frac{\theta_n}{2} - i\delta_{f_j}}
\int \frac{d^4 k}{(2\pi )^4}
\frac{\bar \lambda_n (k\hspace{-.5em}/\, +q\hspace{-.5em}/_2) P_L \nu_i }{\left[ k^2 - m_{\tilde f_{2j}}^2 \right] \left[ (k+q_2)^2 - m_{f_j}^2 \right]}
-( m_{\tilde f_{2j}}^2 \leftrightarrow m_{\tilde f_{1j}}^2)
,
\nonumber\\
\end{eqnarray}
where $q_2$ is the momentum carried by the incident fermion.
The other notations are the same as for Appendix \ref{sec:type1rainbow}.
This contribution does not give the same effective interaction as the renormalizable bilinear RPV interaction due to the Lorentz structure $q\hspace{-.5em}/_2$ of the effective operator,
so it cannot be renormalized away.
It is only generated at the loop level through the trilinear RPV interactions.
After momentum integration, we obtain
\begin{eqnarray}
i{\cal M}_{\nu \chi}
&\approx&
- \frac{i\sqrt{2}}{(4\pi )^2} \hat \lambda_{ijj} n_c (g^{(n)}_{\tilde f_L} + g^{(n)}_{\tilde f_R })
\sin \theta_{f_j} \cos \theta_{f_j} e^{i\frac{\theta_n}{2} - i\delta_{f_j}}
\bar \lambda_n q\hspace{-.5em}/_2 P_L \nu_i 
\int_0^1 dx\,
(1-x)
\ln \left(
\frac{xq_2^2 -m_{\tilde f_{1j}}^2 }{ xq_2^2 -m_{\tilde f_{2j}}^2 }
\right)
\, ,
\label{eq:B2}
\\
i{\cal M}_{\nu^c \chi}
&\approx&
- \frac{i\sqrt{2}}{(4\pi )^2} \hat \lambda^*_{ijj} n_c (g^{(n)}_{\tilde f_L} + g^{(n)}_{\tilde f_R })
\sin \theta_{f_j} \cos \theta_{f_j} e^{i\delta_{f_j} - i\frac{\theta_n}{2} }
\bar \lambda_n q\hspace{-.5em}/_2 P_R \nu_i^c 
\int_0^1 dx\,
(1-x)
\ln \left(
\frac{xq_2^2 -m_{\tilde f_{1j}}^2 }{ xq_2^2 -m_{\tilde f_{2j}}^2 }
\right)
\, ,
\\
i{\cal M}_{\chi \nu}
&\approx&
- \frac{i\sqrt{2}}{(4\pi )^2} \hat \lambda^*_{ijj} n_c (g^{(n)}_{\tilde f_L} + g^{(n)}_{\tilde f_R })
\sin \theta_{f_j} \cos \theta_{f_j} e^{i\delta_{f_j} - i\frac{\theta_n}{2} }
\bar \nu_i q\hspace{-.5em}/_2 P_L \lambda_n 
\int_0^1 dx\,
(1-x)
\ln \left(
\frac{xq_2^2 -m_{\tilde f_{1j}}^2 }{ xq_2^2 -m_{\tilde f_{2j}}^2 }
\right)
\, ,
\label{eq:B4}
\\
i{\cal M}_{\chi \nu^c}
&\approx&
- \frac{i\sqrt{2}}{(4\pi )^2} \hat \lambda_{ijj} n_c (g^{(n)}_{\tilde f_L} + g^{(n)}_{\tilde f_R })
\sin \theta_{f_j} \cos \theta_{f_j} e^{i\frac{\theta_n}{2} - i\delta_{f_j}}
\bar \nu_i q\hspace{-.5em}/_2 P_R \lambda_n 
\int_0^1 dx\,
(1-x)
\ln \left(
\frac{xq_2^2 -m_{\tilde f_{1j}}^2 }{ xq_2^2 -m_{\tilde f_{2j}}^2 }
\right)
\, ,
\end{eqnarray}
where we have also written the off-shell amplitudes given by the transposed diagrams and the contributions from hermitian conjugated interactions.
Here again we have neglected the quark and lepton masses.
By inserting the above effective vertices to the second loop as shown in Fig. \ref{fig:rainbow_type} (b), we obtain the following expression (expanded up to the first order in the momentum $q$ carried by the external gauge boson): 
\begin{eqnarray}
i{\cal M}_{\rm (b)}^{\chi_0}
&\approx&
\frac{2i}{(4\pi )^2} {\rm Im} \Bigl[ \hat \lambda_{ijj} \tilde \lambda^*_{ikk} e^{i(\theta_n -\delta_{f_j})} \Bigr] n_c Q_F e \Bigl(g^{(n)}_{\tilde f_L} + g^{(n)}_{\tilde f_R }\Bigr)
\sin \theta_{f_j} \cos \theta_{f_j} m_{\lambda_n} \epsilon^*_\mu (q)
\nonumber\\
&&\times 
\sum_{\tilde F = \tilde F_L , \tilde F_R}
\int \frac{d^4 k}{(2\pi)^4}
\frac{g^{(n)}_{\tilde F} \bar F_k \gamma_5 F_k \cdot k^2 (-2k^\mu +2p^\mu -q^\mu )}{k^2 \left[ k^2 - m_{\lambda_n}^2  \right] \bigl[ (k+q-p)^2 -m_{\tilde F_k}^2 \bigr] \bigl[ (k-p)^2 -m_{\tilde F_k}^2 \bigr]}
\int_0^1 dx \, (1-x) 
\ln \left(
\frac{xk^2 -m_{\tilde f_{1j}}^2 }{ xk^2 -m_{\tilde f_{2j}}^2 }
\right)
\nonumber\\
&\approx &
\frac{2i}{(4\pi )^2} {\rm Im} \Bigl[\hat \lambda_{ijj} \tilde \lambda^*_{ikk} e^{i(\theta_n -\delta_{f_j})} \Bigr] n_c Q_F e \Bigl(g^{(n)}_{\tilde f_L} + g^{(n)}_{\tilde f_R }\Bigr)
\sin \theta_{f_j} \cos \theta_{f_j} m_{\lambda_n} \epsilon^*_\mu (q)
\nonumber\\
&&\hspace{4em}\times 
\sum_{\tilde F = \tilde F_L , \tilde F_R}
\int \frac{d^4 k}{(2\pi)^4}
\frac{-2g^{(n)}_{\tilde F} m_{\tilde F_k}^2 \bar F_k \gamma_5 F_k p^\mu }{\left[ k^2 - m_{\lambda_n}^2 \right] \bigl[ k^2 -m_{\tilde F_k}^2 \bigr]^3 }
\int_0^1 dx \, (1-x) 
\ln \left(
\frac{xk^2 - m_{\tilde f_{1j}}^2 }{ xk^2 - m_{\tilde f_{2j}}^2 }
\right)
\nonumber\\
&\approx&
\frac{i }{(4\pi )^4} {\rm Im} \Bigl[\hat \lambda_{ijj} \tilde \lambda^*_{ikk} e^{i(\theta_n -\delta_{f_j})} \Bigr] n_c Q_F e \Bigl(g^{(n)}_{\tilde f_L} + g^{(n)}_{\tilde f_R }\Bigr)
\sin \theta_{f_j} \cos \theta_{f_j} m_{\lambda_n} \epsilon^*_\mu (q) \bar F_k \sigma^{\mu \nu} q_\nu \gamma_5 F_k
\nonumber\\
&& \hspace{4em}
\times \sum_{\tilde F= \tilde F_L,\tilde F_R}  g^{(n)}_{\tilde F} m_{\tilde F_k}^2 
\int_0^\infty \hspace{-.5em}
\frac{2 r^2dr }{\bigl[ r +m_{\lambda_n}^2 \bigr] r  \bigl[ r +m_{\tilde F_k}^2 \bigr]^3 } \int_0^1 dx\, (1-x) \ln \left( \frac{xr + m_{\tilde f_{1j}}^2}{xr + m_{\tilde f_{2j}}^2} \right)
\nonumber\\
&=&
\frac{i }{(4\pi )^4} {\rm Im} \Bigl[\hat \lambda_{ijj} \tilde \lambda^*_{ikk} e^{i(\theta_n -\delta_{f_j})} \Bigr] n_c Q_F e \Bigl(g^{(n)}_{\tilde f_L} + g^{(n)}_{\tilde f_R }\Bigr)
\sin \theta_{f_j} \cos \theta_{f_j} m_{\lambda_n} \epsilon^*_\mu (q) \bar F_k \sigma^{\mu \nu} q_\nu \gamma_5 F_k
\nonumber\\
&& \hspace{4em}
\times \sum_{\tilde F= \tilde F_L,\tilde F_R}  g^{(n)}_{\tilde F} m_{\tilde F_k}^2 
\left[ 
G''(m_{\lambda_n}^2 , 0 , m_{\tilde F_k}^2 , m_{\tilde f_{1j}}^2 )
-G''(m_{\lambda_n}^2 , 0 , m_{\tilde F_k}^2 , m_{\tilde f_{2j}}^2 )
\right]
\ ,
\end{eqnarray}
where the function $G''$ is defined in Eq. (\ref{eq:F''}), and $p$ is the external momentum brought by the fermion $F_k$.
Here we only derived the second term of the formula for the EDM (\ref{eq:edmnug}), but the second term of the chargino contribution (\ref{eq:chargino_rainbow})
as well as that of the chromo-EDM (\ref{eq:cedm}) may also be obtained in a similar manner.

\section{\label{sec:type1charginorainbow}Calculation of rainbow diagrams with a chargino}

\begin{figure}[htb]
\begin{center}
\includegraphics[width=9.2cm]{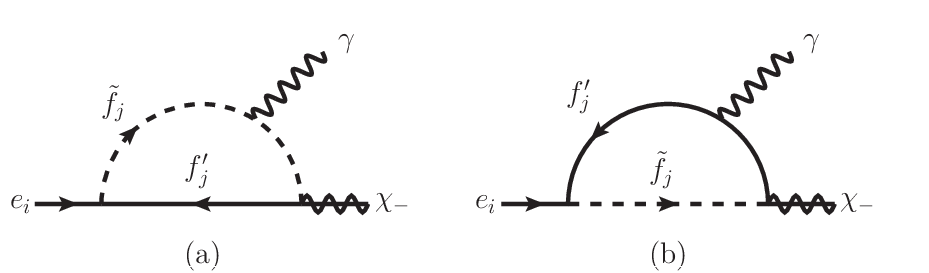}
\caption{\label{fig:egc}
One-loop level diagram contribution to the RPV charged lepton-chargino-photon vertex.
There are less relevant diagrams than Fig. \ref{fig:nugg} due to the projection of chirality by the chargino-fermion-sfermion interaction (\ref{eq:gaugino-int}).
}
\end{center}
\end{figure}

We calculate the rainbow diagrams of Figs. \ref{fig:rainbow_type} with a chargino in the loop.
The off-shell amplitude of the one-loop effective charged lepton-chargino-photon vertex is
\begin{eqnarray}
i{\cal M}_{e \chi \gamma}
&= &
i{\cal M}_{e \chi \gamma}^{\rm (a)}
+i{\cal M}_{e \chi \gamma}^{\rm (b)}
,
\end{eqnarray}
where
\begin{eqnarray}
i{\cal M}_{e \chi \gamma}^{\rm (a)}
&\simeq &
\hat \lambda_{ijj}
\frac{i Q_f e^2 n_c V_{jj}}{(4\pi)^2 \sin \theta_W} \sin \theta_{f_j} \cos \theta_{f_j} e^{-i(\delta_{f_j} -\theta_2 /2 )} \epsilon_\mu^* (q_1) 
\nonumber\\
&& \times
\Biggl[
-\bar {\lambda'_2}^c (q_2) \gamma^\mu P_L e_i (q_1+q_2) 
\int_0^1 \hspace{-0.5em} dz \,
(1-z)\ln \left( z q_2^2 - m_{\tilde f_{2j}}^2 \right) 
-2\bar {\lambda'_2}^c (q_2) q\hspace{-.5em}/_2 P_L e_i (q_1+q_2) q_2^\mu
\int_0^1 \hspace{-0.5em} dz
\frac{z(1-z) }{ z q_2^2 - m_{\tilde f_{2j}}^2 } 
\Biggr]
\nonumber\\
&&
-( m_{\tilde f_{2j}}^2 \leftrightarrow m_{\tilde f_{1j}}^2) 
\ ,
\label{eq:echigamma(a)}
\end{eqnarray}
and
\begin{eqnarray}
i{\cal M}_{e \chi \gamma}^{\rm (b)}
&\simeq &
\hat \lambda_{ijj}
\frac{i Q_{f'} e^2 n_c V_{jj} }{(4\pi)^2\sin \theta_W} \sin \theta_{f_j} \cos \theta_{f_j} e^{-i(\delta_{f_j} -\theta_2 /2 )} \epsilon_\mu^* (q_1) 
\nonumber\\
&& \times
\Biggl[
\bar {\lambda'_2}^c (q_2) \gamma^\mu P_L 
e_i
(q_1+q_2)
\int_0^1 \hspace{-0.5em} dz \, 
z \ln \left( z q_2^2 - m_{\tilde f_{2j}}^2 \right)
\nonumber\\
&& \hspace{1.5em}
+ \bar {\lambda'_2}^c (q_2) 
\int_0^1 \hspace{-0.5em} dz \,
\left\{
\frac{
-z q\hspace{-.5em}/_2 q\hspace{-.5em}/_1 \gamma^\mu 
+ z (1-z)(2q_2^\mu q\hspace{-.5em}/_2 -q_2^2 \gamma^\mu )
}{z q_2^2 - m_{\tilde f_{2j}}^2 } 
\right\} P_L 
e_i
(q_1+q_2)
\Biggr]
-( m_{\tilde f_{2j}}^2 \leftrightarrow m_{\tilde f_{1j}}^2)
.\ \ \ \ 
\label{eq:echigamma(b)}
\end{eqnarray}
Here we only kept terms contributing to the EDM and neglected quark and lepton masses.
The above amplitudes, $i{\cal M}_{e \chi \gamma}^{\rm (a)}$ and $i{\cal M}_{e \chi \gamma}^{\rm (b)}$, correspond to the diagrams of Figs. \ref{fig:egc} (a) and (b), respectively.
We note that $f,F$ ($\tilde f,\tilde F$) are down-type quarks (squarks) or charged leptons (sleptons), while $f',F'$ ($\tilde f',\tilde F'$) are up-type quarks (squarks) or neutrinos (sneutrinos), respecting the definition of $\hat \lambda$ and $\tilde \lambda$ as defined below Eq. (\ref{eq:RPV4f}).
The matrix $V$ is the Cabibbo-Kobayashi-Maskawa matrix for quark-squark loops, otherwise it is a unit matrix.
The sum of Eqs. (\ref{eq:echigamma(a)}) and (\ref{eq:echigamma(b)}) then yields
\begin{eqnarray}
i{\cal M}_{e \chi \gamma}
&\simeq &
\hat \lambda_{ijj}
\frac{i e^2 n_c V_{jj} }{(4\pi)^2\sin \theta_W} \sin \theta_{f_j} \cos \theta_{f_j} e^{-i(\delta_{f_j} -\theta_2 /2 )} \epsilon_\mu^* (q_1) 
\nonumber\\
&& \times
\int_0^1 \hspace{-0.5em} dz \, 
(Q_f- Q_{f'} )
\bar {\lambda'_2}^c (q_2) 
\Biggl[
(1-z) \ln \left( z q_2^2 - m_{\tilde f_{1j}}^2 \right)
\gamma^\mu 
+ \frac{2 z (1-z) q_2^\mu q\hspace{-.5em}/_2 }{z q_2^2 - m_{\tilde f_{1j}}^2 } 
\Biggr]
P_L e_i (q_1+q_2)
\nonumber\\
&&
-( m_{\tilde f_{1j}}^2 \leftrightarrow m_{\tilde f_{2j}}^2)
.\ \ \ \ 
\end{eqnarray}
In this derivation, we used the formula (\ref{eq:01integralformula}).
As for Eq. (\ref{eq:m(a)+m(b)}), we have omitted Lorentz structures such as $q_2^\mu q\hspace{-.5em}/_1 $ and $(q_1 \cdot q_2) \gamma^\mu $ since they do not contribute to the final EDM.
The above expression is consistent with Eq. (\ref{eq:m(a)+m(b)}) if we take $Q_f = Q_{f'}$ which is the case for the diagrams with neutralino and neutrino (see Appendix \ref{sec:type1rainbow}).
Similarly, the transposed diagrams yield
\begin{eqnarray}
i{\cal M}_{\chi e \gamma}
&\simeq &
\hat \lambda_{ijj}^*
\frac{i e^2 n_c V_{jj} }{(4\pi)^2\sin \theta_W} \sin \theta_{f_j} \cos \theta_{f_j} e^{i(\delta_{f_j} -\theta_2 /2 )} \epsilon_\mu^* (q_1) 
\nonumber\\
&& \times
\int_0^1 \hspace{-0.5em} dz \, 
\bar e_i (q_2) 
\Biggl\{
(Q_f- Q_{f'} )
\Biggl[
(1-z) \ln \left( z q_2^2 - m_{\tilde f_{1j}}^2 \right)
\gamma^\mu 
+ \frac{2 z (1-z) q_2^\mu q\hspace{-.5em}/_2 }{z q_2^2 - m_{\tilde f_{1j}}^2 } 
\Biggr]
\nonumber\\
&& \hspace{7em}
+ Q_{f'} 
\frac{
z q\hspace{-.5em}/_2 q\hspace{-.5em}/_1 \gamma^\mu 
}{z q_2^2 - m_{\tilde f_{1j}}^2 } 
\Biggr\}
P_L {\lambda'_2}^c (q_1+q_2)
-( m_{\tilde f_{1j}}^2 \leftrightarrow m_{\tilde f_{2j}}^2)
.\ \ \ \ 
\end{eqnarray}

We now insert the effective vertices $i{\cal M}_{e \chi \gamma}$ and $i{\cal M}_{\chi e \gamma}$ into the second loop [Fig. \ref{fig:rainbow_type} (a)] and obtain the following expression: 
\begin{eqnarray}
i{\cal M}_{\rm (a)}^{\chi_-}
&\simeq &
-\hat \lambda_{ijj} \tilde \lambda_{ikk}^* e^{-i(\delta_{f_j} -\theta_2 )} \frac{e^3 n_c V_{jj} V_{kk} }{(4\pi)^2 \sin^2 \theta_W} \sin \theta_{f_j} \cos \theta_{f_j} m_{\lambda_2}
\epsilon_\mu^* (q)
\nonumber\\
&& \hspace{1em} \times
\int \frac{d^4 k}{(2\pi )^4} 
\frac{1}{\Bigl[ k^2 - m_{\lambda_2}^2 \Bigr] \Bigl[ (k+q)^2 - m_{e_i}^2 \Bigr] \Bigl[ (k+q-p)^2 - m_{\tilde F'_{Lk}}^2 \Bigr]}
\nonumber\\
&&\hspace{4em} \times 
\Biggl\{
( Q_f - Q_{f'} )
\bar F_k (p-q) \gamma^\mu (k\hspace{-.5em}/\, +q\hspace{-.5em}/\, ) P_R F_k (p)
\int_0^1 \hspace{-.5em} dx\, (1-x) \ln \left( xk^2 -m_{\tilde f_{1j}}^2 \right)
\nonumber\\
&&\hspace{5.5em} 
+2 ( Q_f - Q_{f'} )
\bar F_k (p-q) (
k^2
+k\hspace{-.5em}/\, q\hspace{-.5em}/\, ) k^\mu P_R F_k (p)
\int_0^1 \hspace{-.5em} dx  \frac{x(1-x)}{ xk^2 -m_{\tilde f_{1j}}^2}
\ \ \ 
\Biggr\}
\nonumber\\
&&
-\hat \lambda_{ijj}^* \tilde \lambda_{ikk} e^{i(\delta_{f_j} -\theta_2 )} \frac{e^3 n_c V_{jj} V_{kk} }{(4\pi)^2 \sin^2 \theta_W} \sin \theta_{f_j} \cos \theta_{f_j} m_{\lambda_2}
\epsilon_\mu^* (q) 
\nonumber\\
&& \hspace{1em} \times
\int \frac{d^4 k}{(2\pi )^4} 
\frac{1}{\Bigl[ (k+q)^2 - m_{\lambda_2}^2 \Bigr] \Bigl[ k^2 - m_{e_i}^2 \Bigr] \Bigl[ (k+q-p)^2 - m_{\tilde F'_{Lk}}^2 \Bigr]}
\nonumber\\
&&\hspace{4em} \times 
\Biggl\{
( Q_f - Q_{f'} )
\bar F_k (p-q) k\hspace{-.5em}/\, \gamma^\mu P_L F_k (p)
\int_0^1 \hspace{-.5em} dx\, (1-x) \ln \left( xk^2 -m_{\tilde f_{1j}}^2 \right)
\nonumber\\
&&\hspace{5.5em} 
+2 ( Q_f - Q_{f'} )
k^2 k^\mu \bar F_k (p-q) P_L F_k (p)
\int_0^1 \hspace{-.5em} dx  \frac{x(1-x)}{ xk^2 -m_{\tilde f_{1j}}^2}
\nonumber\\
&&\hspace{5.5em} 
+ Q_{f'} k^2 \bar F_k (p-q) q\hspace{-.5em}/\, \gamma^\mu P_L F_k (p)
\int_0^1 \hspace{-.5em} dx  \frac{x}{ xk^2 -m_{\tilde f_{1j}}^2}
\ \ \ 
\Biggr\}
\nonumber\\
&&
-(m_{\tilde f_{1j}}^2 \leftrightarrow m_{\tilde f_{2j}}^2 )
.
\label{eq:M(a)chi-}
\end{eqnarray}

Let us now calculate the rainbow diagrams with the external photon attached to the charged lepton of the second loop [see Fig. \ref{fig:rainbow_type} (c)].
For that we need the one-loop level effective charged lepton-chargino transition 
\begin{eqnarray}
i{\cal M}_{e \chi}
&\approx &
\hat \lambda_{ijj} \frac{ i e n_c V_{jj} }{(4\pi )^2 \sin \theta_W}  \sin \theta_{f_j} \cos \theta_{f_j} e^{-i(\delta_{f_j} -\theta_2 /2 )} 
\bar {\lambda'_2}^c (q_2) \, q\hspace{-.5em}/_2 P_L e_i (q_2)
\int_0^1 dx\,
(1-x)
\ln \left(
\frac{xq_2^2 -m_{\tilde f_{1j}}^2 }{ xq_2^2 -m_{\tilde f_{2j}}^2 }
\right)
,
\\
i{\cal M}_{\chi e}
&\approx &
\hat \lambda_{ijj}^* \frac{ i e n_c V_{jj} }{(4\pi )^2 \sin \theta_W}  \sin \theta_{f_j} \cos \theta_{f_j} e^{i(\delta_{f_j} -\theta_2 /2 )} 
\bar e_i (q_2) \, q\hspace{-.5em}/_2 P_L {\lambda'_2}^c (q_2)
\int_0^1 dx\,
(1-x)
\ln \left(
\frac{xq_2^2 -m_{\tilde f_{1j}}^2 }{ xq_2^2 -m_{\tilde f_{2j}}^2 }
\right)
.
\end{eqnarray}
These can be derived in the same way as Eqs. (\ref{eq:B2}) and (\ref{eq:B4}) (see Appendix \ref{sec:type2rainbow}), by just replacing coupling constants and mass parameters.
After inserting them into the second loop, we obtain
\begin{eqnarray}
i{\cal M}_{\rm (c)}^{\chi_-}
&\approx &
-\hat \lambda_{ijj} \tilde \lambda_{ikk}^* \frac{e^3 n_c V_{jj} V_{kk} }{(4\pi )^2 \sin^2 \theta_W } \sin \theta_{f_j} \cos \theta_{f_j} e^{-i(\delta_{f_j} -\theta_2 )} m_{\lambda_2} \epsilon^*_\mu (q)
\nonumber\\
&&
\times \int \frac{d^4 k}{(2\pi)^4}
\frac{\bar F_k (p-q) 
k^2
\gamma^\mu (k\hspace{-.5em}/\, + q\hspace{-.5em}/\, 
) P_R F_k(p)}{\Bigl[ k^2 - m_{\lambda_2}^2 \Bigr] \Bigl[ k^2 - m_{e_i}^2 \Bigr] \Bigl[ (k+q)^2 - m_{e_i}^2 \Bigr] \Bigl[ (k+q-p)^2 - m_{\tilde F'_{Lk}}^2 \Bigr]}
\int_0^1 \hspace{-.5em} dx (1-x) \ln \left( \frac{xk^2 -m_{\tilde f_{1j}}^2 }{ xk^2 -m_{\tilde f_{2j}}^2 } \right)
\nonumber\\
&& -
\hat \lambda_{ijj}^* \tilde \lambda_{ikk} \frac{e^3 n_c V_{jj} V_{kk} }{(4\pi )^2 \sin^2 \theta_W } \sin \theta_{f_j} \cos \theta_{f_j} e^{i(\delta_{f_j} -\theta_2 )} m_{\lambda_2} \epsilon^*_\mu (q)
\nonumber\\
&&
\times \int \frac{d^4 k}{(2\pi)^4}
\frac{\bar F_k (p-q) k\hspace{-.5em}/\, 
\gamma^\mu (k + q )^2 P_L F_k(p)}{\Bigl[ (k+q)^2 - m_{\lambda_2}^2 \Bigr] \Bigl[ (k+q)^2 - m_{e_i}^2 \Bigr] \Bigl[ k^2 - m_{e_i}^2 \Bigr] \Bigl[ (k+q-p)^2 - m_{\tilde F'_{Lk}}^2 \Bigr]}
\nonumber\\
&& \hspace{4em} \times
\int_0^1 \hspace{-.5em} dx (1-x) \ln \left( \frac{x(k+q)^2 -m_{\tilde f_{1j}}^2 }{ x(k+q)^2 -m_{\tilde f_{2j}}^2 } \right)
\nonumber\\
&\approx &
-\hat \lambda_{ijj} \tilde \lambda_{ikk}^* \frac{e^3 n_c V_{jj} V_{kk} }{(4\pi )^2 \sin^2 \theta_W} \sin \theta_{f_j} \cos \theta_{f_j} e^{-i(\delta_{f_j} -\theta_2 )} m_{\lambda_2} \epsilon^*_\mu (q)
\nonumber\\
&&
\times \int \frac{d^4 k}{(2\pi)^4}
\frac{
\bar F_k (p-q) 
\gamma^\mu ( k\hspace{-.5em}/\, +q\hspace{-.5em}/\, ) 
P_R F_k(p)
}{\Bigl[ k^2 - m_{\lambda_2}^2 \Bigr] \Bigl[ (k+q)^2 - m_{e_i}^2 \Bigr] \Bigl[ (k+q-p)^2 - m_{\tilde F'_{Lk}}^2 \Bigr]}
\int_0^1 \hspace{-.5em} dx (1-x) \ln \left( \frac{xk^2 -m_{\tilde f_{1j}}^2 }{ xk^2 -m_{\tilde f_{2j}}^2 } \right)
\nonumber\\
&& -
\hat \lambda_{ijj}^* \tilde \lambda_{ikk} \frac{e^3 n_c V_{jj} V_{kk} }{(4\pi )^2 \sin^2 \theta_W } \sin \theta_{f_j} \cos \theta_{f_j} e^{i(\delta_{f_j} -\theta_2 )} m_{\lambda_2} \epsilon^*_\mu (q)
\nonumber\\
&&
\times \int \frac{d^4 k}{(2\pi)^4}
\frac{\bar F_k (p-q) k\hspace{-.5em}/\, 
\gamma^\mu P_L F_k(p)}{\Bigl[ (k+q)^2 - m_{\lambda_2}^2 \Bigr] \Bigl[ k^2 - m_{e_i}^2 \Bigr] \Bigl[ (k+q-p)^2 - m_{\tilde F'_{Lk}}^2 \Bigr]}
\int_0^1 \hspace{-.5em} dx (1-x) \ln \left( \frac{xk^2 -m_{\tilde f_{1j}}^2 }{ xk^2 -m_{\tilde f_{2j}}^2 } \right)
\nonumber\\
&& -
\hat \lambda_{ijj}^* \tilde \lambda_{ikk} \frac{e^3 n_c V_{jj} V_{kk} }{(4\pi )^2 \sin^2 \theta_W } \sin \theta_{f_j} \cos \theta_{f_j} e^{i(\delta_{f_j} -\theta_2 )} m_{\lambda_2} \epsilon^*_\mu (q)
\nonumber\\
&&
\times \int \frac{d^4 k}{(2\pi)^4}
\frac{\bar F_k (p-q) k\hspace{-.5em}/\, 
\gamma^\mu P_L F_k(p)}{\Bigl[ k^2 - m_{\lambda_2}^2 \Bigr] \Bigl[ k^2 - m_{e_i}^2 \Bigr] \Bigl[ k^2 - m_{\tilde F'_{Lk}}^2 \Bigr]}
\int_0^1 \hspace{-.5em} dx 
\Biggl[
\frac{ 2 x(1-x) k\cdot q }{xk^2 -m_{\tilde f_{1j}}^2}
-\frac{ 2 x(1-x) k\cdot q }{xk^2 -m_{\tilde f_{2j}}^2}
\Biggr]
.\ \ \ 
\label{eq:(u2)}
\end{eqnarray}
The last term of the second equality comes from the first order expansion of the logarithm in $q$.
We see that the two logarithmic terms of the last equality cancel with those of Eq. (\ref{eq:M(a)chi-}).

We also give the expression for the rainbow diagrams with the external photon attached to the chargino of the second loop [see Fig. \ref{fig:rainbow_type} (d)]:
\begin{eqnarray}
i{\cal M}_{\rm (d)}^{\chi_-}
&\approx &
-\hat \lambda_{ijj} \tilde \lambda_{ikk}^* \frac{e^3 n_c V_{jj} V_{kk} }{(4\pi )^2 \sin^2 \theta_W} \sin \theta_{f_j} \cos \theta_{f_j} e^{-i(\delta_{f_j} -\theta_2 )} m_{\lambda_2} \epsilon^*_\mu (q)
\nonumber\\
&&
\hspace{1em} \times \int \frac{d^4 k}{(2\pi)^4}
\frac{\bar F_k (p-q) [\gamma^\mu (k\hspace{-.5em}/\, +q\hspace{-.5em}/\, ) + k\hspace{-.5em}/\, \gamma^\mu ] P_R F_k (p) (k+q)^2}{\Bigl[ (k+q)^2 - m_{\lambda_2}^2 \Bigr] \Bigl[ k^2 - m_{\lambda_2}^2 \Bigr] \Bigl[ (k+q)^2 - m_{e_i}^2 \Bigr] \Bigl[ (k+q-p)^2 - m_{\tilde F'_{Lk}}^2 \Bigr]}
\nonumber\\
&& \hspace{6em} \times
\int_0^1 \hspace{-.5em} dx (1-x) \ln \left( \frac{x(k+q)^2 -m_{\tilde f_{1j}}^2 }{ x(k+q)^2 -m_{\tilde f_{2j}}^2 } \right)
\nonumber\\
&&
-\hat \lambda_{ijj}^* \tilde \lambda_{ikk} \frac{e^3 n_c V_{jj} V_{kk} }{(4\pi )^2 \sin^2 \theta_W} \sin \theta_{f_j} \cos \theta_{f_j} e^{i(\delta_{f_j} -\theta_2 )} m_{\lambda_2} \epsilon^*_\mu (q)
\nonumber\\
&&
\hspace{1em} \times \int \frac{d^4 k}{(2\pi)^4}
\frac{\bar F_k (p-q) [\gamma^\mu (k\hspace{-.5em}/\, +q\hspace{-.5em}/\, ) + k\hspace{-.5em}/\, \gamma^\mu ] P_L F_k (p) k^2}{\Bigl[ (k+q)^2 - m_{\lambda_2}^2 \Bigr] \Bigl[ k^2 - m_{\lambda_2}^2 \Bigr] \Bigl[ k^2 - m_{e_i}^2 \Bigr] \Bigl[ (k+q-p)^2 - m_{\tilde F'_{Lk}}^2 \Bigr]}
\nonumber\\
&& \hspace{6em} \times
\int_0^1 \hspace{-.5em} dx (1-x) \ln \left( \frac{xk^2 -m_{\tilde f_{1j}}^2 }{ xk^2 -m_{\tilde f_{2j}}^2 } \right)
\nonumber\\
&\approx &
-\hat \lambda_{ijj} \tilde \lambda_{ikk}^* \frac{e^3 n_c V_{jj} V_{kk} }{(4\pi )^2 \sin^2 \theta_W} \sin \theta_{f_j} \cos \theta_{f_j} e^{-i(\delta_{f_j} -\theta_2 )} m_{\lambda_2} \epsilon^*_\mu (q)
\nonumber\\
&&
\hspace{1em} \times \int \frac{d^4 k}{(2\pi)^4}
\frac{k^2}{ \Bigl[ k^2 - m_{\lambda_2}^2 \Bigr]^2 \Bigl[ k^2 - m_{e_i}^2 \Bigr] \Bigl[ k^2 - m_{\tilde F'_{Lk}}^2 \Bigr]}
\int_0^1 \hspace{-.5em} dx (1-x) \ln \left( \frac{xk^2 -m_{\tilde f_{1j}}^2 }{ xk^2 -m_{\tilde f_{2j}}^2 } \right)
\nonumber\\
&&\hspace{5em} \times 
\Biggl[
\bar F_k (p-q) \gamma^\mu q\hspace{-.5em}/\, P_R F_k (p)
+\frac{\bar F_k (p-q) p^\mu P_R F_k (p) \cdot k^2}{k^2 -m_{\tilde F'_{Lk}}^2 }
\Biggr]
\nonumber\\
&&
-\hat \lambda_{ijj}^* \tilde \lambda_{ikk} \frac{e^3 n_c V_{jj} V_{kk} }{(4\pi )^2 \sin^2 \theta_W} \sin \theta_{f_j} \cos \theta_{f_j} e^{i(\delta_{f_j} -\theta_2 )} m_{\lambda_2} \epsilon^*_\mu (q)
\nonumber\\
&&
\hspace{1em} \times \int \frac{d^4 k}{(2\pi)^4}
\frac{k^2}{ \Bigl[ k^2 - m_{\lambda_2}^2 \Bigr]^2 \Bigl[ k^2 - m_{e_i}^2 \Bigr] \Bigl[ k^2 - m_{\tilde F'_{Lk}}^2 \Bigr]}
\int_0^1 \hspace{-.5em} dx (1-x) \ln \left( \frac{xk^2 -m_{\tilde f_{1j}}^2 }{ xk^2 -m_{\tilde f_{2j}}^2 } \right)
\nonumber\\
&&\hspace{5em} \times 
\Biggl[
\bar F_k (p-q) \gamma^\mu q\hspace{-.5em}/\, P_L F_k (p)
+\frac{\bar F_k (p-q) p^\mu P_L F_k (p) \cdot k^2}{k^2 -m_{\tilde F'_{Lk}}^2 }
\Biggr]
\nonumber\\
&\simeq &
i\, {\rm Im} \Bigl[ \hat \lambda_{ijj} \tilde \lambda_{ikk}^* e^{-i(\delta_{f_j} -\theta_2 )} \Bigr] \frac{e^3 n_c V_{jj} V_{kk} }{2 (4\pi )^2 \sin^2 \theta_W} \sin \theta_{f_j} \cos \theta_{f_j} m_{\lambda_2} \epsilon^*_\mu (q) \bar F i\sigma^{\mu \nu} q_\nu \gamma_5 F
\nonumber\\
&&
\times \int \frac{d^4 k}{(2\pi)^4}
\frac{k^2}{ \Bigl[ k^2 - m_{\lambda_2}^2 \Bigr]^2 \Bigl[ k^2 - m_{e_i}^2 \Bigr] \Bigl[ k^2 - m_{\tilde F'_{Lk}}^2 \Bigr]}
\Biggl[
1
- \frac{m_{\tilde F'_{Lk}}^2}{k^2 -m_{\tilde F'_{Lk}}^2 }
\Biggr]
\int_0^1 \hspace{-.5em} dx (1-x) \ln \left( \frac{xk^2 -m_{\tilde f_{1j}}^2 }{ xk^2 -m_{\tilde f_{2j}}^2 } \right)
,\ \ \ \ \ \ 
\end{eqnarray}
where $\simeq$ means that we only kept Lorentz structures contributing to the EDM operator.
In this derivation we have used
\begin{eqnarray}
\bar F \gamma^\mu q\hspace{-.5em}/\, \gamma_5 F
&\simeq &
-i \bar F \sigma^{\mu \nu} q_\nu \gamma_5 F \, ,
\\
\bar F p^\mu \gamma_5 F 
&\simeq &
\frac{i}{2} \bar F \sigma^{\mu \nu} q_\nu \gamma_5 F 
\, .
\end{eqnarray}

Finally, the diagram of Fig. \ref{fig:rainbow_type} (b) can be obtained in the same way as that of $i{\cal M}_{\rm (b)}^{\chi_0}$ derived in Appendix \ref{sec:type2rainbow}.
By just replacing coupling constants and mass parameters, we have
\begin{eqnarray}
i{\cal M}_{\rm (b)}^{\chi_-}
&=&
{\rm Im} \Bigl[\hat \lambda_{ijj} \tilde \lambda^*_{ikk} e^{i(\theta_2 -\delta_{f_j})} \Bigr] \frac{n_c Q_{F'} e^3 V_{jj} V_{kk} }{2(4\pi )^4 \sin^2 \theta_W} 
\sin \theta_{f_j} \cos \theta_{f_j} m_{\lambda_2} \epsilon^*_\mu (q) \bar F_k i \sigma^{\mu \nu} q_\nu \gamma_5 F_k
\nonumber\\
&& \hspace{4em}
\times 
m_{\tilde F'_{Lk}}^2 
\left[ 
G''(m_{\lambda_2}^2 , m_{e_i}^2 , m_{\tilde F'_{Lk}}^2 , m_{\tilde f_{1j}}^2 )
-G''(m_{\lambda_2}^2 , m_{e_i}^2 , m_{\tilde F'_{Lk}}^2 , m_{\tilde f_{2j}}^2 )
\right]
.
\end{eqnarray}

The overall sum is
\begin{eqnarray}
i{\cal M}_{F_k}^{\chi_-}
&\approx&
-
{\rm Im} \bigl[ \hat \lambda_{ijj} \tilde \lambda^*_{ikk} e^{i(\theta_2 -\delta_{f_j} )} \bigr]
\frac{ e^3 n_c V_{jj} V_{kk} }{2\sin^2 \theta_W (4\pi )^4 }  \sin \theta_{f_j} \cos \theta_{f_j} m_{\lambda_2} \epsilon^*_\mu (q) \bar F_k i\sigma^{\mu \nu} q_\nu \gamma_5 F_k
\nonumber\\
&& \times
\int_0^\infty \hspace{-.5em}d r
\Biggl\{
Q_{f'} \frac{r^2}{\Bigl[ r +m_{\lambda_2}^2 \Bigr] \Bigl[ r+ m_{e_i}^2 \Bigr] \Bigl[ r +m_{\tilde F'_{Lk}}^2 \Bigr] } \int_0^1 \hspace{-0.5em}dx \frac{x}{xr + m_{\tilde f_{1j}}^2} 
\nonumber\\
&&\hspace{5em}
+ \frac{m_{\tilde F'_{Lk}}^2 r}{\Bigl[ r + m_{\lambda_2}^2 \Bigr] \Bigl[ r + m_{e_i}^2 \Bigr] \Bigl[ r + m_{\tilde F'_{Lk}}^2 \Bigr]^2 } \int_0^1 \hspace{-0.5em}dx\, (2x-1 )\ln (xr + m_{\tilde f_{1j}}^2) 
\nonumber\\
&&\hspace{5em}
- \frac{r^2}{\Bigl[ r + m_{\lambda_2}^2 \Bigr]^2 \Bigl[ r + m_{e_i}^2 \Bigr] \Bigl[ r + m_{\tilde F'_{Lk}}^2 \Bigr] } \int_0^1 \hspace{-0.5em}dx (1-x)\ln (xr + m_{\tilde f_{1j}}^2) 
\nonumber\\
&&\hspace{5em}
- \frac{m_{\tilde F'_{Lk}}^2 r^2 }{\Bigl[ r + m_{\lambda_2}^2 \Bigr]^2 \Bigl[ r + m_{e_i}^2 \Bigr] \Bigl[ r + m_{\tilde F'_{Lk}}^2 \Bigr]^2 } \int_0^1 \hspace{-0.5em}dx (1-x)\ln (xr + m_{\tilde f_{1j}}^2) 
\ \ \Biggr\}
-(m_{\tilde f_{1j}}^2 \leftrightarrow m_{\tilde f_{2j}}^2)
\nonumber\\
&&
+i{\cal M}_{\rm (b)}^{\chi_-}
\nonumber\\
&\approx&
-
{\rm Im} \bigl[ \hat \lambda_{ijj} \tilde \lambda^*_{ikk} e^{i(\theta_2 -\delta_{f_j} )} \bigr]
\frac{ e^3 n_c V_{jj} V_{kk} }{2\sin^2 \theta_W (4\pi )^4 }  \sin \theta_{f_j} \cos \theta_{f_j} m_{\lambda_2} \epsilon^*_\mu (q) \bar F_k i\sigma^{\mu \nu} q_\nu \gamma_5 F_k
\nonumber\\
&& \times
\int_0^\infty \hspace{-.5em}d r
\Biggl\{
Q_{f'} \frac{r^2}{\Bigl[ r +m_{\lambda_2}^2 \Bigr] \Bigl[ r+ m_{e_i}^2 \Bigr] \Bigl[ r +m_{\tilde F'_{Lk}}^2 \Bigr] } \int_0^1 \hspace{-0.5em}dx \frac{x}{xr + m_{\tilde f_{1j}}^2} 
\nonumber\\
&&\hspace{5em}
+ \frac{m_{\tilde F'_{Lk}}^2 r}{\Bigl[ r + m_{\lambda_2}^2 \Bigr] \Bigl[ r + m_{e_i}^2 \Bigr] \Bigl[ r + m_{\tilde F'_{Lk}}^2 \Bigr]^2 } \int_0^1 \hspace{-0.5em}dx\, \ln (xr + m_{\tilde f_{1j}}^2) 
\nonumber\\
&&\hspace{5em}
- \frac{r^2}{\Bigl[ r + m_{\lambda_2}^2 \Bigr]^2 \Bigl[ r + m_{e_i}^2 \Bigr] \Bigl[ r + m_{\tilde F'_{Lk}}^2 \Bigr] } \int_0^1 \hspace{-0.5em}dx (1-x)\ln (xr + m_{\tilde f_{1j}}^2) 
\nonumber\\
&&\hspace{5em}
- \frac{m_{\tilde F'_{Lk}}^2 (3r^3 +2m_{\lambda_2}^2r^2) }{\Bigl[ r + m_{\lambda_2}^2 \Bigr]^2 \Bigl[ r + m_{e_i}^2 \Bigr]^2 \Bigl[ r + m_{\tilde F'_{Lk}}^2 \Bigr]^2 } \int_0^1 \hspace{-0.5em}dx (1-x)\ln (xr + m_{\tilde f_{1j}}^2) 
\ \ \Biggr\}
-(m_{\tilde f_{1j}}^2 \leftrightarrow m_{\tilde f_{2j}}^2)
\nonumber\\
&&
+i{\cal M}_{\rm (b)}^{\chi_-}
. \ \ \ \ \ 
\end{eqnarray}
Let us show the analytic forms of the relevant integrals.
The first term in the curly bracket has the same form as that of $i{\cal M}^{\chi_0}_{\rm (a)}$, derived in Appendix \ref{sec:type1rainbow}.
The integral of the first term is
\begin{eqnarray}
\int_0^\infty \hspace{-.5em}
\frac{r^2dr }{( r +a )( r+ b )( r +c ) } \int_0^1 \hspace{-0.5em} \frac{xdx}{xr + d_1} -(d_1 \leftrightarrow d_2)
&=&
G' (a,b,c,d_1 ) 
-G' (a,b,c,d_2 ) 
\, .
\end{eqnarray}
The function $G'$ is defined in Eq. (\ref{eq:F'}).

To calculate the other terms, we need to evaluate the following integrals:
\begin{eqnarray}
&&
\int_0^\infty \hspace{-0.5em} dr\,
\frac{r}{(r+a)(r+b)(r+c)^2} \int_0^1 \hspace{-0.5em} dx\, \ln \left( \frac{xr+d_1}{xr+d_2} \right)
\, ,
\label{eq:intergal1}
\\
&&
\int_0^\infty \hspace{-0.5em} dr\,
\frac{r^2}{(r+a)(r+b)(r+c)^2} \int_0^1 \hspace{-0.5em} dx\, (1-x) \ln \left( \frac{xr+d_1}{xr+d_2} \right)
\, ,
\label{eq:intergal2}
\\
&&
\int_0^\infty \hspace{-0.5em} dr\,
\frac{r^3}{(r+a)^2(r+b)^2(r+c)^2} \int_0^1 \hspace{-0.5em} dx\, (1-x) \ln \left( \frac{xr+d_1}{xr+d_2} \right)
\, ,
\label{eq:intergal3}
\\
&&
\int_0^\infty \hspace{-0.5em} dr\,
\frac{r^2}{(r+a)^2(r+b)^2(r+c)^2} \int_0^1 \hspace{-0.5em} dx\, (1-x) \ln \left( \frac{xr+d_1}{xr+d_2} \right)
\, .
\label{eq:intergal4}
\end{eqnarray}

The integral of Eq. (\ref{eq:intergal1}) is
\begin{eqnarray}
&&
\int_0^\infty \hspace{-0.5em} dr\,
\frac{r}{(r+a)(r+b)(r+c)^2} \int_0^1 \hspace{-0.5em} dx\, \ln \left( \frac{xr+d_1}{xr+d_2} \right)
\, ,
\nonumber\\
&=&
\frac{a -d_1}{(a-b)(a-c)^2} \left[ {\rm Li}_2 \left( 1-\frac{a}{d_1} \right) -{\rm Li}_2 \left( 1-\frac{c}{d_1} \right) \right] 
+\frac{b-d_1}{(b-a)(b-c)^2} \left[ {\rm Li}_2 \left( 1-\frac{b}{d_1} \right) -{\rm Li}_2 \left( 1-\frac{c}{d_1} \right) \right] 
\nonumber\\
&&
+\frac{a }{(a-b)(a-c)^2} \ln d_1 \ln \frac{c}{a} 
+\frac{b}{(b-a)(b-c)^2} \ln d_1 \ln \frac{c}{b} 
\nonumber\\
&&
-(d_1 \leftrightarrow d_2 )
\nonumber\\
&=&
G^{c_1} (a,b,c,d_1) 
-G^{c_1} (a,b,c,d_2) 
\, .
\label{eq:doubleintegral1}
\end{eqnarray}
The function $G^{c_1}$ is defined in Eq. (\ref{eq:fc1}).

The next integral [Eq. (\ref{eq:intergal2})] can be transformed as
\begin{eqnarray}
&&
\int_0^\infty \hspace{-0.5em} dr\,
\frac{r^2}{(r+a)(r+b)(r+c)^2} \int_0^1 \hspace{-0.5em} dx\, (1-x) \ln \left( \frac{xr+d_1}{xr+d_2} \right)
\nonumber\\
&=&
-\frac{(a-d_1)^2}{2(a-b)(a-c)^2} \left[ {\rm Li}_2 \left( 1-\frac{a}{d_1} \right) -{\rm Li}_2 \left( 1-\frac{c}{d_1} \right) \right] 
-\frac{(b-d_1)^2}{2(b-a)(b-c)^2} \left[ {\rm Li}_2 \left( 1-\frac{b}{d_1} \right) -{\rm Li}_2 \left( 1-\frac{c}{d_1} \right) \right] 
\nonumber\\
&& 
+\frac{(c-d_1) \ln c + d_1 (1+ \ln d_1)}{2(c-a)(c-b)} 
-\frac{ad_1 + a^2 \ln d_1}{2(a-b)(a-c)^2} \ln \frac{c}{a} 
-\frac{bd_1 + b^2 \ln d_1}{2(b-a)(b-c)^2} \ln \frac{c}{b} 
\nonumber\\
&&
-(d_1 \leftrightarrow d_2 )
\nonumber\\
&=&
G^{c_2} (a,b,c,d_1) 
-G^{c_2} (a,b,c,d_2) 
\, .
\end{eqnarray}
The function $G^{c_2}$ is defined in Eq. (\ref{eq:fc2}).

The third integral [Eq. (\ref{eq:intergal3})] can be transformed as
\begin{eqnarray}
&&
\int_0^\infty \hspace{-0.5em} dr\,
\frac{r^3}{(r+a)^2(r+b)^2(r+c)^2} \int_0^1 \hspace{-0.5em} dx\, (1-x) \ln \left( \frac{xr+d_1}{xr+d_2} \right)
\nonumber\\
&=&
\frac{2a^2 (a^2-bc) - a^2 (a-b)(a-c) }{2(a-b)^3(a-c)^3} \ln a \ln d_1 
+\frac{a d_1}{2(a-b)^2(a-c)^2} \left[ \ln \left( \frac{a}{d_1} \right) -1 \right]
\nonumber\\
&&
+
\frac{ad_1 (a^2 -bc)}{(a-b)^3(a-c)^3} \ln a 
-
\frac{2(d_1-a)^2 (a^2 -bc) +(d_1^2-a^2)(a-b)(a-c)}{2(a-b)^3(a-c)^3} {\rm Li}_2 \left( 1-\frac{a}{d_1} \right) 
\nonumber\\
&&
+(\mbox{even permutations of } a, b, c)
\nonumber\\
&&
-(d_1 \leftrightarrow d_2 )
\nonumber\\
&=&
G^{c_3} (a,b,c,d_1) 
-G^{c_3} (a,b,c,d_2) 
\, .
\end{eqnarray}
The function $G^{c_3}$ is defined in Eq. (\ref{eq:fc3}).

The final integral [Eq. (\ref{eq:intergal4})] can be transformed as
\begin{eqnarray}
&&
\int_0^\infty \hspace{-0.5em} dr\,
\frac{r^2}{(r+a)^2(r+b)^2(r+c)^2} \int_0^1 \hspace{-0.5em} dx\, (1-x) \ln \left( \frac{xr+d_1}{xr+d_2} \right)
\nonumber\\
&=&
\frac{(a -d_1)(a^2-bc-2ad_1+bd_1+cd_1)}{(a-b)^3(a-c)^3} {\rm Li}_2 \left( 1-\frac{a}{d_1} \right) 
-\frac{a (a^2-bc)}{(a-b)^3(a-c)^3} \ln a \ln d_1
+\frac{d_1 (1+\ln d_1)}{2(a-b)^2(a-c)^2}
\nonumber\\
&&
+\frac{a d_1 ( -2a+b+c)}{(a-b)^3(a-c)^3} \ln a
\nonumber\\
&&
+(\mbox{even permutations of } a, b, c)
\nonumber\\
&&
-(d_1 \leftrightarrow d_2 )
\nonumber\\
&=&
G^{c_4} (a,b,c,d_1) 
-G^{c_4} (a,b,c,d_2) 
\, .
\end{eqnarray}
The function $G^{c_4}$ is defined in Eq. (\ref{eq:fc4}).
Combining everything, we find Eq. (\ref{eq:chargino_rainbow}).

\twocolumngrid

\end{document}